\documentclass[letter]{elsarticle}

\usepackage{url}
\usepackage[english]{babel} \usepackage{graphicx} \usepackage{amsmath}\usepackage{amsthm}
\usepackage{amssymb} \usepackage{proof} \usepackage{bbm}

\newcommand{\pb}{{\mathbf{pb}}}
\newcommand{\pf}{{\mathbf{pf}}}
\newcommand{\bb}{{\mathbf{bb}}}
\newcommand{\byf}{{\mathbf{bf}}}

\newcommand{\msg}{\mathcal{M}}
\newcommand{\processes}{\mathcal{I}}
 \newcommand{\boldS}{{\mathbf{S}}}
\newcommand{\lang}{{\mathcal{L}}} \newcommand{\slicegraphvertex}{\mathfrak{v}}
 \newcommand{\canonical}{\mathcal{C}}

\newcommand{\N}{{\mathbbm{N}}} 
 
 \newcommand{\hasse}{{\mathcal{H}}}
 \newcommand{\transS}{\mathcal{T}}
\newcommand{\partition}{\mathcal{P}} 
 
\newcommand{\automaton}{{\mathcal{A}}} \newcommand{\slice}{{\mathcal{S}}}
\newcommand{\sliceautomaton}{\mathcal{S}\!\mathcal{A}}

\newcommand{\slicegraph}{\mathcal{S}\!\mathcal{G}}
\newcommand{\slicealphabet}{\Sigma_{\mathbb{S}}}

\newcommand{\hasseGenerator}{\hasse\!\mathcal{G}}
\newcommand{\takesp}{{\hat{p}}} \newcommand{\putsp}{{\check{p}}}
\newtheorem{theorem}{Theorem}
\newtheorem{proposition}{Proposition}
\newtheorem{lemma}{Lemma}
\newtheorem{corollary}{Corollary}

\newtheorem{definition}{Definition}

\begin{document}

\title{Canonizable Partial Order Generators \\
	and Regular Slice Languages{\small \footnote{This work extends the paper \cite{deOliveiraOliveira2012} by the same author.}}}
\author{Mateus de Oliveira Oliveira}
\address{School of Computer Science and Communication, \\ KTH Royal Institute of Technology, 
100-44 Stockholm, Sweden\\ mdeoliv@kth.se}

\begin{abstract} 
In a previous work we introduced slice graphs as a way to specify both infinite
languages of directed acyclic graphs (DAGs) and infinite languages of partial
orders. Therein we focused on the study of Hasse diagram generators, i.e.,
slice graphs that generate only transitive reduced DAGs. In the present work 
we show that any slice graph can be transitive reduced into a Hasse diagram generator
representing the same set of partial orders. By employing this result we establish 
unknown connections between the true concurrent behavior of bounded $p/t$-nets 
and traditional approaches for representing infinite families of partial orders, 
such as Mazurkiewicz trace languages and Message Sequence Chart ($MSC$) languages. Going further, 
we identify the family of weakly saturated slice graphs. The class of partial order languages 
that can be represented by weakly saturated slice graphs is closed under union, intersection and 
even under a suitable notion of complementation (globally bounded complementation). The partial order
languages in this class also admit canonical representatives in terms of Hasse diagram generators, 
and have decidable inclusion and emptiness of intersection. Our transitive reduction algorithm 
plays a fundamental role in these decidability results. 

\end{abstract}

\begin{keyword}
Partial Order Languages \sep 
Regular Slice Languages \sep \\ \hspace{1.8cm} Transitive Reduction \sep
Petri Nets
\end{keyword}

\maketitle

\section{Introduction} \label{introduction} 

It is widely recognized that both the true concurrency and the causality
between the events of concurrent systems can be adequately captured through
partial orders
\cite{Gischer1988,GaifmanPratt1987,Vogler1992,JagadeesanJagadeesan1995,LangerakBrinksmaKatoen1997}.
In order to represent the whole concurrent behavior of systems, several methods
of specifying infinite families of partial orders have been proposed. Partial
languages \cite{Gabowski1982}, series-parallel languages \cite{LodayaWeil2000},
concurrent automata \cite{Droste1992}, causal automata
\cite{MontanariPistore1997}, approaches derived from trace theory
\cite{FranchonMorin2009,Mazurkiewicz1986,Diekert1994,HoogersKleijnThiagarajan1995,KuskeMorin2002},
approaches derived from message sequence chart theory
\cite{HenriksenMukundKumarSohoniThiagarajan2005,GazagnaireGenestHelouet2007,GenestMuschollSeidlZeitoun2006},
and more recently, Hasse diagram generators \cite{deOliveiraOliveira2010}.  

Hasse diagram generators are defined with basis on slice graphs, which 
by their turn, may be regarded as a specialization (modulo some convenient notational
adaptations) of graph grammars \cite{EngelfrietVereijken1997,Courcelle1987}.
Indeed, slice graphs may be viewed as automata that concatenate atomic blocks
called slices, to generate infinite families of directed acyclic graphs (DAGs)
and to represent infinite sets of partial orders. A Hasse diagram generator
$\hasseGenerator$ is a slice graph that generates exclusively transitive
reduced graphs. In other words, every DAG in the graph language generated by
$\hasseGenerator$ is the Hasse diagram of the partial order it represents.
Such generators were introduced by us in \cite{deOliveiraOliveira2010}
in the context of Petri net theory, and used to solve different open problems related to the partial order
semantics of bounded $place/transition$-nets ($p/t$-nets). 
For instance, we showed that the set of partial order runs of any bounded $p/t$-net $N$ can be represented by an effectively 
constructible Hasse diagram generator $\hasseGenerator_N$.
Previously, approaches that mapped behavioral objects to $p/t$-nets were
either not expressive enough to fully capture partial order behavior 
of bounded $p/t$-nets, or were not guaranteed to be finite
and thus, not effective \cite{EsparzaRomerVogler2002,McMillan1992,HoogersKleijnThiagarajan1996,HaymanWinskel2008}. 

In \cite{deOliveiraOliveira2010} we also showed how to use Hasse diagram generators
to verify the partial order behavior of concurrent systems modeled through bounded $p/t$-nets. 
More precisely, given a bounded $p/t$-net $N$
with partial order behavior $\lang_{PO}(N)$ and a HDG $\hasseGenerator$
representing a set $\lang_{PO}(\hasseGenerator)$ of partial orders, we may
effectively verify both whether $\lang_{PO}(\hasseGenerator)$ is included into
$\lang_{PO}(\hasseGenerator)$ and whether their intersection is empty.
Previously an analogous verification result was only known for finite languages
of partial orders \cite{JuhasLorenzDesel2005}. As a meta-application of this
verification result, we were able to test the inclusion of the partial order
behavior of two bounded $p/t$-nets $N_1$ and $N_2$: Compute 
$\hasseGenerator_{N_1}$ and test whether 
$\lang_{PO}(\hasseGenerator_{N_1})\subseteq \lang_{PO}(N_2)$.  The possibility
of performing such an inclusion test for bounded $p/t$-nets had been open for
at least a decade. In the nineties, Jategaonkar-Jagadeesan and Meyer
\cite{JategaonkarMeyer1996} proved that the inclusion of the causal behavior of
$1$-safe $p/t$-nets is decidable, and Montanari and Pistore
\cite{MontanariPistore1997} showed how to determine whether two bounded nets
have bisimilar causal behaviors. 

Finally, Hasse diagram generators may be used to address the synthesis of
concurrent systems from behavioral specifications. The idea of the synthesis is
appealing: Instead of constructing a system and verifying if it behaves as
expected, we specify a priori which runs should be present on it, and then
automatically construct a system satisfying the given specification
\cite{KupfermanVardiYannakakis2011,PnueliRosner1989,Church1962}. In our setting
the systems are modeled via $p/t$-nets and the specification is made in terms
of Hasse diagram generators. In \cite{deOliveiraOliveira2010} we devised an
algorithm that takes a Hasse diagram generator $\hasseGenerator$ and a bound
$b$ as input, and determines whether there is a $b$-bounded $p/t$-net whose
partial order behavior includes $\lang_{PO}(\hasseGenerator)$. If such a net
exists, the algorithm returns the net $N$ whose behavior minimally includes
$\lang_{PO}(\hasseGenerator)$.  More precisely for every other $b$-bounded
$p/t$-net $N'$ satisfying $\lang_{PO}(\hasseGenerator)\subseteq \lang_{PO}(N')$
it is guaranteed that $\lang_{PO}(N)\subseteq \lang_{PO}(N')$. This implies in
particular, that if the set of runs specified by $\hasseGenerator$ indeed
matches the partial order behavior of a $b$-bounded $p/t$-net $N$, then this
net will be returned.  The synthesis of $p/t$-nets from finite sets of partial
orders was accomplished in \cite{BergenthumDeselLorenzMauserFundamenta2008} and
subsequently generalized in \cite{BergenthumDeselLorenzMauserInfinite2008} (see also \cite{Mauser2010}) to
infinite languages specified by rational expressions over partial orders, which
are nevertheless not expressive enough to represent the whole behavior of
arbitrary bounded $p/t$-nets. For other results considering the synthesis of
several types of Petri nets from several types of automata and languages,
specifying both sequential and step behaviors we point to
\cite{EhrenfeuchtRozenberg1989a,HoogersKleijnThiagarajan1995,BadouelDarondeau1996a,BadouelDarondeau1998,Darondeau1998,Darondeau2000}. 

\section{Transitive Reduction of Slice Graphs and its Consequences}
\label{section:Consequences}

Both the verification and the synthesis of $p/t$-nets described in the previous section are stated in function 
of Hasse diagram generators, and do not extend directly to general slice graphs. The main goal of this paper 
is to overcome this limitation, by proving that any slice graph can be transitive reduced into a Hasse diagram 
generator specifying the same partial order language.

\paragraph{{\bf Theorem \ref{theorem:TransitiveReductionSliceGraphsB} (Transitive Reduction of General Slice Graphs)}}
Any slice graph $\slicegraph$ can be transitive reduced into a Hasse diagram generator 
$\hasseGenerator$ representing the same partial order language, i.e., $\lang_{PO}(\slicegraph) = \lang_{PO}(\hasseGenerator)$.\\

This result is interesting for two main reasons: First slice graphs are much more flexible than 
Hasse diagram generators from a specification point of view. Second it establishes interesting connections 
between $p/t$-nets and well known formalisms aimed to specify infinite families of partial orders, such as Mazurkiewicz trace languages 
\cite{Mazurkiewicz1986} and message sequence chart (MSC) languages \cite{HenriksenMukundKumarSohoniThiagarajan2005}. 
More precisely, we prove that if a partial order language $\lang_{PO}$ is specified 
through a pair $(\automaton,I)$ of finite automaton $\automaton$ over an alphabet of events $\Sigma$ and a Mazurkiewicz 
independence relation $I\subseteq \Sigma\times\Sigma$, then there is a slice graph $\slicegraph$ 
representing the same set of partial orders. A similar result holds if $\lang_{PO}$ is specified by a high-level
message sequence chart (HMSC), or equivalently, by a message sequence graph (MSG) \cite{AlurYannakakis1999,MuschollPeledSu1998,Morin2001}.  
We point out that in general, the slice graphs arising from these transformations may be far from being transitive reduced and that a direct 
translation of these approaches in terms of Hasse diagram generators is not evident. Nevertheless, 
Theorem \ref{theorem:TransitiveReductionSliceGraphsB} guarantees that these slice graphs can be indeed transitive reduced into Hasse diagram 
generators representing the same partial order language, allowing us in this way to apply both our verification and synthesis results 
to Mazurkiewicz trace languages and MSC languages (Corollary \ref{corollary:MSCAndPetriNets}). 

It is worth noting that Corollary \ref{corollary:MSCAndPetriNets} addresses the synthesis of {\em unlabeled} $p/t$-nets from partial order languages represented by traces or message sequence graphs. 
The synthesis of labeled $p/t$-nets (i.e., nets in which two transitions may be labeled by the same action) from Mazurkiewicz trace 
languages and from local trace languages \cite{HoogersKleijnThiagarajan1996} was addressed respectively in \cite{HussonMorin2000} and in \cite{KuskeMorin2002}. 
However there is a substantial difference between labeled and unlabeled $p/t$-nets when it comes to partial order behavior. 
For instance, if we allow the synthesized nets to be labeled, we are helped by the fact that labeled $1$-safe $p/t$-nets are already 
as partial order expressive as their $b$-bounded counterparts \cite{BestWimmel2000}. Thus the synthesis of unlabeled nets tends to be harder. 

Our transitive reduction algorithm is also a necessary step towards the canonization of slice graphs. We say that a function 
$\canonical_{PO}$ canonizes slice graphs with respect to their {\bf partial 
order} languages if for every slice graph $\slicegraph$,
$\lang_{PO}(\slicegraph) =\lang_{PO}(\canonical_{PO}(\slicegraph))$ and $\canonical_{PO}(\slicegraph)\sim\canonical_{PO}(\slicegraph')$ for all 
other slice graph $\slicegraph'$ satisfying $\lang_{PO}(\slicegraph) = \lang_{PO}(\slicegraph')$. 
In the same way that a Hasse diagram provides a minimal representation for its induced partial order, it is natural 
that Hasse diagram generators correspond to the canonical forms of slice graphs. 
However simply transitive reducing 
a slice graph is not sufficient to put it into a canonical form, and indeed canonization is in general uncomputable. 
Fortunately, there is a very natural and decidable\footnote{In \cite{HenriksenMukundKumarSohoniThiagarajan2005} it is undecidable
whether a MSC-language is linearization-regular. This is not in contradiction with the decidability of weak saturation. An analogous statement for us 
would be: It is undecidable whether a slice graph can be weakly saturated.} subclass of slice graphs (weakly saturated slice graphs) for which canonization is feasible.
Besides admitting canonical representatives, 
partial order languages represented by weakly saturated slice graphs are closed under union, intersection and even under 
a special notion of complementation, which we call {\em globally bounded complementation}. Furthermore inclusion (and consequently, equality) and 
emptiness of intersection are decidable for this class of languages. Transitive reduction will play an important role in the definition of 
globally bounded complementation and, as we argue in the next paragraph, it will play a {\bf fundamental role} in the closure, decidability 
and canonizability results stated above. 

A slice graph $\slicegraph$ is meant to represent three distinct languages: A slice language $\lang(\slicegraph)$ which 
is a regular subset of the free monoid generated by a slice alphabet $\slicealphabet^c$; 
a graph language $\lang_G(\slicegraph)$ consisting of the DAGs which have a string representative in the slice language; and a partial order language 
$\lang_{PO}(\slicegraph)$ obtained by taking the transitive closure of DAGs in the graph language. As we will show in Section \ref{section:Saturation},
any weakly saturated slice graph can be efficiently transformed into a stronger form, which we call {\em saturated slice graph}, representing the 
same graph and partial order languages. It turns out that except for complementation,  
operations involving languages of DAGs generated by saturated slice graphs are reflected by operations performed in their slice languages, which 
for being regular, have several well known decidability and computability results. This observation may be interpreted as a consequence of the fact 
that saturated slice languages are closed under a certain commutation operation defined on $\slicealphabet^c$. If additionally, 
the slice graphs in consideration are Hasse diagram generators,
then questions about their partial order languages can be further mapped to questions about their graph languages, paving in this way a path to decidability. 
The crucial point is that this last observation fails badly if the slice graphs are not transitive reduced: There exist
({\bf even saturated}) slice graphs $\slicegraph$ and $\slicegraph'$ for which $\lang_{G}(\slicegraph) \cap \lang_{G}(\slicegraph') = \emptyset$ but 
$\lang_{PO}(\slicegraph) \cap \lang_{PO}(\slicegraph') \neq \emptyset$, or for which $\lang_{G}(\slicegraph)\nsubseteq \lang_{G}(\slicegraph')$ but 
$\lang_{PO}(\slicegraph)\subseteq \lang_{PO}(\slicegraph')$. Thus it is essential that we transitive reduce slice graphs before 
performing operations with their partial order languages. With regard to this observation, an important feature of our transitive 
reduction algorithm is that it preserves weak saturation. The complementation of the graph and of the partial order languages generated by  
a saturated slice graphs  does not follows from the closure under commutation described above, however it is still achievable in a suitable 
sense (globally bounded complementation), whose definition we postpone to Section \ref{section:Saturation}. 

A skeptic could wonder whether weak saturation is an excessively strong condition which could be only satisfied 
by uninteresting examples of slice graphs. We counter this skepticism by describing three natural situations 
in which weakly saturated slice graphs arise: The first two examples stem from the fact that our 
study of weakly saturated slice languages was inspired, and indeed generalizes, both the theory of recognizable trace 
languages \cite{Mazurkiewicz1986} and the theory of linearization-regular\footnote{In our work the term regular is used in the standard sense of finite automata theory. 
The notion of "regular" used in \cite{HenriksenMukundKumarSohoniThiagarajan2005} is analogous to our notion of regular+saturated. } 
message sequence languages \cite{HenriksenMukundKumarSohoniThiagarajan2005}. In particular, recognizable trace languages can be
mapped to weakly saturated regular slice languages, while linearization-regular MSC languages which are representable by message sequence graphs, 
may be mapped to {\em loop connected} slice graphs, which can be efficiently weakly saturated. Our third and most important
example comes from the theory of bounded $p/t$-nets. More precisely, we show that 
the Hasse diagram generators associated to bounded $p/t$-nets in \cite{deOliveiraOliveira2010} are saturated. 
 This last observation has two important consequences: first, slice graphs are strictly more expressive than 
both Mazurkiewicz trace languages, and MSC-languages, since there exist even $1$-safe $p/t$-nets whose partial order behavior 
cannot be expressed through these formalisms; second, it implies that the behavior of bounded $p/t$-nets may be canonically represented by Hasse diagram generators. 
While in \cite{deOliveiraOliveira2010} we were able to associate a HDG $\hasseGenerator_N$ to any bounded $p/t$-net 
$N$, we were not able to prove that if two nets $N$ and $N'$ have the same partial order behavior then they can be associated 
to same HDG\footnote{In general, a partial order language can be represented by several distinct Hasse diagram generators.}. 
By showing that the partial order language of bounded $p/t$-nets may be represented via saturated slice languages
we are able to achieve precisely this goal (Theorem \ref{theorem:RefinedExpressibility}): 

\paragraph{{\bf Theorem \ref{theorem:RefinedExpressibility} (Intuitive Version) }}
The set $\lang_{PO}(N)$ of partial order runs of any bounded $p/t$-net $N$ can be canonically represented by a saturated 
Hasse diagram generator $\hasseGenerator(N)$. In particular for any other bounded $p/t$-net $N$ such that 
$\lang_{PO}(N)=\lang_{PO}(N')$ it holds that $\hasseGenerator(N)=\hasseGenerator(N')$. \\ 

The rest of the paper is organized as follows: 
Next, in Section \ref{section:Slices} we define slices,
slice graphs and slice languages. Subsequently, in sections \ref{section:TransitiveReduction} and  \ref{section:Saturation} 
we introduce the main contributions of this work, which are our transitive reduction algorithm (Section \ref{section:TransitiveReduction})
and our study of partial order languages that can be represented trough saturated slice languages (Section \ref{section:Saturation}). 
In section \ref{section:Reductions} we prove that both Mazurkiewicz trace languages and 
MSC-languages can be mapped to slice languages. In section \ref{section:PetriNets} we show 
how our results may be used as a link between Mazurkiewicz traces, MSC languages 
and the partial order behavior of $p/t$-nets. Finally in 
Section \ref{section:FinalComments} we make some final comments. 

\section{Slices} \label{section:Slices}

There are several automata-theoretic approaches for the specification of
infinite families of graphs: graph automata
\cite{Thomas1992,BrandenburgSkodinis2005}, automata over planar DAGs
\cite{BossutDauchetWarin1995}, graph rewriting systems
\cite{Courcelle1987,BauderonCourcelle1987,EngelfrietVereijken1997}, and others
\cite{GiammarresiRestivo1996,GiammarresiRestivo1992,BozapalidisKalampakas2006}.
In this section we will introduce an approach that is more suitable for our
needs. Namely, the representation of infinite families of DAGs with bounded
slice width. In particular, the slices defined in this section can be regarded as
a specialized version of the multi-pointed graphs defined in
\cite{EngelfrietVereijken1997}, which are too general, and which are subject to
a slightly different notion of concatenation.  

\begin{figure}[hf] \centering \includegraphics[scale=0.35]{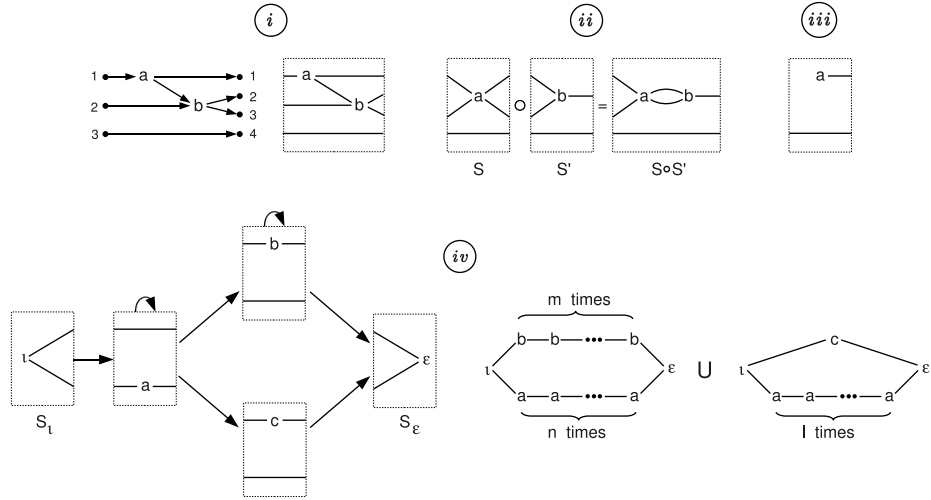}
\caption{ $i$) A slice $ii$) Composition of slices. $iii$) A degenerate slice.
$iv$) A slice graph labeled with unit slices, and an intuitive representation
of its graph language. $\boldS_{\iota}$ is initial and $\boldS_{\varepsilon}$,
final.} \label{figure:sliceBasics} \end{figure}

A {\em slice} is a labeled DAG $\boldS=(V,E,l)$ whose vertex set $V$ is
partitioned into three subsets: A non-empty center $C$ labeled by $l$ with the
elements of an arbitrary set $T$ of events, and the in- and out-frontiers $I$
and $O$ respectively which are numbered by $l$ in such a way that $l(I)=\{1,
\cdots,|I|\}$ and $l(O)=\{1,...,|O|\}$.  Furthermore a unique edge in $E$
touches each frontier vertex $v\in I\dot{\cup} O$, where $\dot{\cup}$ denotes
the disjoint union of sets. This edge is outgoing if $v$ lies on the
in-frontier $I$ and incoming if $v$ lies on the out-frontier $O$.
In drawings, we surround slices by dashed rectangles, and implicitly direct
their edges from left to right.  In-frontier and out-frontier vertices are
determined respectively by the intersection of edges with the left and right
sides of the rectangle. Frontier vertices are implicitly numbered from top to
bottom. Center vertices are indicated by their labels (Fig.
\ref{figure:sliceBasics}-$i$).

A slice $\boldS_1$ can be composed with a slice $\boldS_2$ whenever the
out-frontier of $\boldS_1$ is of the same size as the in-frontier of
$\boldS_2$.  In this case, the resulting slice $\boldS_1\circ \boldS_2$ is
obtained by gluing the single edge touching the $j$-th out-frontier vertex of
$\boldS_1$ to the corresponding edge touching the $j$-th in-frontier vertex of
$\boldS_2$ (Fig. \ref{figure:sliceBasics}-$ii$).  We note that as a result of
the composition, multiple edges may arise, since the vertices on the glued
frontiers disappear.  A slice with a unique vertex in the center is called a
{\em unit slice}. A sequence of unit slices $\boldS_1\boldS_2 \cdots \boldS_n$
is a {\em unit decomposition} of a slice $\boldS$ if $\boldS =
\boldS_1\circ\boldS_2\circ \cdots \circ \boldS_n$. The definition of unit
decomposition extends to DAGs by regarding them as slices with empty in and
out-frontiers. The slice-width of a slice is defined as the size of its greatest
frontier. The slice width of a unit decomposition $\boldS=
\boldS_1\circ\boldS_2\circ \cdots \circ \boldS_n$ is the slice-width of its
widest slice. The {\em existential slice-width} of a DAG $G$ is the slice width 
of its thinest unit decomposition. The {\em global slice-width} of a DAG $G$ is the width 
of its widest unit decomposition.

We say that a slice is {\em initial} if its in-frontier is empty and {\em
final} if its out-frontier is empty.  A unit slice is non-degenerate if its
center vertex is connected to at least one in-frontier (out-frontier) vertex
whenever the in-frontier (out)-frontier is not empty. In Fig.
\ref{figure:sliceBasics}-$iii$ we depict a degenerate unit slice. 
A {\em slice alphabet} is any finite set $\slicealphabet$ of slices. The slice alphabet 
of width $c$ over a set of events $T$ is the set $\slicealphabet^c$ of all unit slices of 
width at most $c$, whose center vertex is labeled with an event from $T$. A {\em slice language} over a slice alphabet $\slicealphabet$ is a subset $\lang\subseteq
\slicealphabet^*$ where for each string $\boldS_1\boldS_2\cdots\boldS_n\in
\lang$, $\boldS_1$ is initial, $\boldS_n$ is final and $\boldS_i$ can be
composed with $\boldS_{i+1}$ for $1\leq i < n$. From a slice language $\lang$
we may derive a language $\lang_G$ of DAGs by composing the slices in the
strings of $\lang$, and a language $\lang_{PO}$ of partial orders, by taking
the transitive closure of each DAG in $\lang_G$:

\begin{equation} \label{equation:graphLanguage} \lang_G =
\{\boldS_1\circ\boldS_2\circ \cdots\circ \boldS_n | \boldS_1\boldS_2\cdots
\boldS_n \in \lang\} \hspace{0.5cm} \mbox{ and } \hspace{0.5cm}
\lang_{PO}=\{H^*| H\in \lang_{G}\} \end{equation}

In this paper we assume that all slices in a slice alphabet $\slicealphabet$
are unit and non-degenerate, but this restriction is not crucial. With this
assumption however, every DAG in the graph language derived from a slice
language has a unique minimal and a unique maximal vertex.

A slice language is regular if it is generated by a finite automaton or by a regular
expression over slices\footnote{The operation of the monoid is just the
concatenation $\boldS_1\boldS_2$ of slice symbols $\boldS_1$ and $\boldS_2$ and
should not be confused with the composition $\boldS_1 \circ \boldS_2$ of
slices.}. We notice that a slice language is a subset of the free monoid
generated by a slice alphabet $\slicealphabet$ and thus we do not need to make
a distinction between regular and rational slice languages. In particular every
slice language generated by a regular expression can be also generated by a
finite automaton. Equivalently, a slice language is regular if and only if it
can be generated by the slice graphs defined below
\cite{deOliveiraOliveira2010}: 

\begin{definition}[Slice Graph] A {\em slice graph} over a slice alphabet
$\slicealphabet$ is a labeled directed graph
$\slicegraph=(\mathcal{V},\mathcal{E},\mathcal{S})$ possibly containing loops
but without multiple edges. The function $\mathcal{S}:\mathcal{V}\rightarrow
\Sigma_{\mathbb{S}}$ satisfies the following condition: $(v_1,v_2) \in
\mathcal{E}$ implies that $\mathcal{S}(v_1)$ can be composed with
$\mathcal{S}(v_2)$.  We say that a vertex on a slice graph is initial if it is
labeled with an initial slice and final if it is labeled with a final slice. We
denote $\lang(\slicegraph)$ the slice language generated by $\slicegraph$,
which we define as:
$$\lang(\slicegraph)=\{\mathcal{S}(v_1)\mathcal{S}(v_2)\cdots\mathcal{S}(v_n):
v_1v_2\cdots v_n \mbox{ is a walk on $\slicegraph$ from an initial to a final
vertex} \}$$ \end{definition}

We write respectively $\lang_G(\slicegraph)$ and $\lang_{PO}(\slicegraph)$ for
the graph and the partial order languages derived from $\lang(\slicegraph)$. A
slice language $\lang$ is {\em transitive reduced} if all DAGs in $\lang_G$ are
simple and transitive reduced. In other words, each DAG in $\lang_G$ is the
Hasse diagram of a partial order in $\lang_{PO}$. A slice graph is a {\em Hasse
diagram generator} if its slice language is transitive reduced.

%%%%%%%%%%%%%%%%%%%%%%%%%%%%%%%%%%%%%%%%%%%%%%%%%%%%%%%%%%%%%%%%%%%%%%%%%%%%%%

%%%%%%%%%%%%%%%%%%%%%%%%%%%%%%%%%%%%%%%%%%%%%%%%%%%%%%%%%%%%%%%%%%%%%%%%%%%%%%

\section{Sliced Transitive Reduction} \label{section:TransitiveReduction}

In \cite{deOliveiraOliveira2010} we devised a method to filter out from the
graph language of a slice graph $\slicegraph$ all $DAGs$ which are not
transitive reduced. In this way we were able to obtain a Hasse diagram
generator $\hasseGenerator$ whose graph language consists precisely on the
Hasse diagrams generated by $\slicegraph$ (i.e. $\lang_{PO}(\hasseGenerator)\subseteq \lang_{PO}(\slicegraph)$).  The method
we devised therein falls short of being a transitive reduction algorithm, since
the partial order generated by the resulting slice graph $\hasseGenerator$
could be significantly shrunk and indeed even reduced to the empty set. It was
not even clear whether such a task could be accomplished at all, since we are
dealing with applying a non-trivial algorithm, i.e. the transitive reduction,
to an infinite number of DAGs at the same time. Fortunately in this section we prove
that such a transitive reduction is accomplishable, by developing an algorithm that
takes a slice graph as input and returns a Hasse diagram generator $\hasseGenerator$
satisfying $\lang_{PO}(\slicegraph)=\lang_{PO}(\hasseGenerator)$.

\begin{figure}[h] \centering
\includegraphics[scale=0.35]{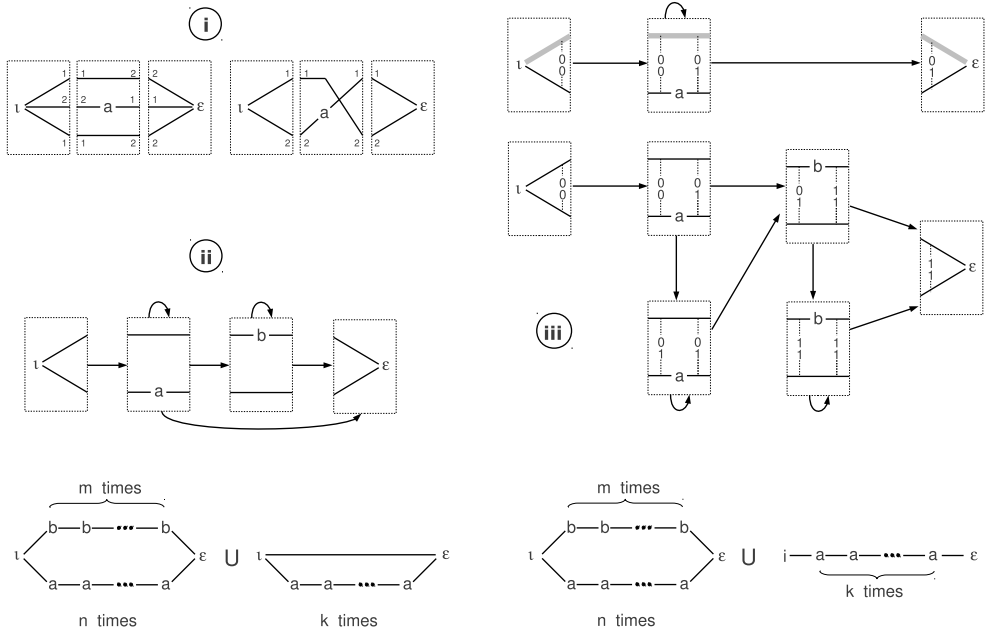}
\caption{ $i$) An illustration of how multiple edges are collapsed in the proof
of Theorem \ref{theorem:TransitiveReductionSliceGraphsB}. $ii$) A slice graph
and an intuitive depiction of its graph language. $iii)$ The Hasse diagram
generator obtained from the slice graph to the left. The marked edges (in
gray), which should be deleted, and the values of the coloring were left to
illustrate the proof of Theorem \ref{theorem:TransitiveReductionSliceGraphsA}.
} \label{figure:TransitiveReductionSliceGraphs} \end{figure}

The difficulty in devising an algorithm to transitive reduce slice graphs stems
from the fact that a slice that labels a vertex $\mathfrak{v}$ of a slice graph
may be used to form both $DAGs$ which are transitive reduced and $DAGs$ which
are not, depending on which path we are considering in the slice graph.  This
observation is illustrated in Figure
\ref{figure:TransitiveReductionSliceGraphs}.$ii$ where the slice containing the
event $a$ has this property. Thus in general the transitive reduction cannot be
performed independently on each slice of the slice graph.  To overcome 
 this difficulty we will introduce in Definition
\ref{definition:TransitivitySeasoning} and in Lemma
\ref{lemma:SlicedTransitiveReduction} a "sliced" characterization of
superfluous edges of DAGs, i.e., edges that do not carry any useful
transitivity information.  By expanding each slice of the slice graph with a set
of specially tagged copies satisfying the conditions listed in Definition
\ref{definition:TransitivitySeasoning} and connecting them in a special way, we
will be able to keep all paths which give rise to transitive reduced DAGs, and to
create new paths which will give rise to transitive reduced versions of the
non-transitive DAGs generated by the original slice graph. 
In the proof of Theorem \ref{theorem:TransitiveReductionSliceGraphsA} we develop an algorithm that
transitive reduces slice graphs which do not generate DAGs with multiple edges.
Subsequently, in Theorem \ref{theorem:TransitiveReductionSliceGraphsB} we will
eliminate the restriction on multiple edges and prove that slice graphs in
general can be transitive reduced.

\begin{figure}[h] \centering
\includegraphics[scale=0.45]{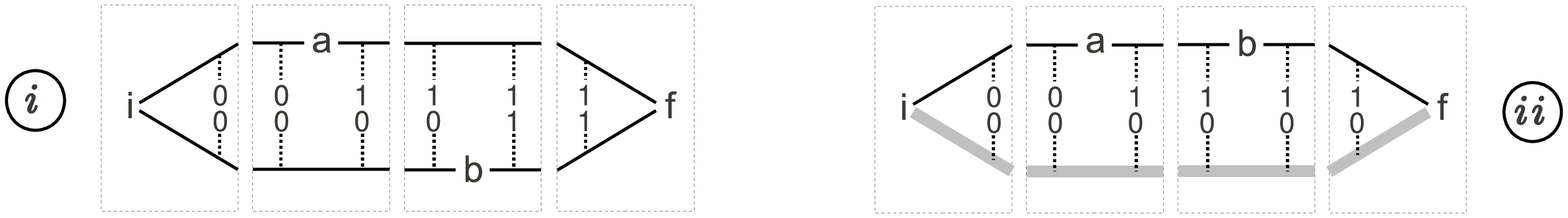}
\caption{ The transitivity coloring of the unit decomposition of two DAGs.
$i$) The DAG is transitive reduced. No edge is marked. $ii$) The DAG is not
transitive reduced.  The sliced parts of each superfluous edge are marked (in
gray). Deleting the marked edges and composing the slices we are left with the
transitive reduction of the original DAG.} \label{figure:HasseEvaluationAlone}
\end{figure}

We say that an edge $e$ of a simple DAG $H$ is superfluous if the transitive
closure of $H$ equals the transitive closure of $H\backslash \{e\}$.  In this
section we will develop a method to highlight the sliced parts of superfluous
edges of a graph $H$ on any of its unit decompositions
$\boldS_1\boldS_2\cdots\boldS_n$ (Fig. \ref{figure:HasseEvaluationAlone}-$vi$).
Deleting these highlighted edges from each slice of the decomposition, we are
left with a unit decomposition $\boldS_1'\boldS_2'\cdots\boldS_n'$ of the
transitive reduction of $H$. It turns out that we may transpose this process to
slice graphs. Thus given a slice graph $\slicegraph$ we will be able to
effectively compute a Hasse diagram generator $\hasseGenerator$ that represents
the same language of partial order as $\slicegraph$, i.e.
$\lang_{PO}(\slicegraph)=\lang_{PO}(\hasseGenerator)$. 

A function $\transS:E^2\rightarrow\{0,1\}^2$ defined on the edges of a unit
slice $\boldS=(\{v\},E,l)$ is called a coloring of $\boldS$. A sequence of
functions $\transS_1\transS_2\cdots\transS_n$ is a coloring of a unit
decomposition $\boldS_1\boldS_2 \cdots \boldS_n$ of a DAG $H$ if each
$\transS_i$ is a coloring of $\boldS_i$ and if the colors associated by
$\transS_i$ to pairs of edges touching the out-frontier of $\boldS_i$ agree
with the colors  associated by $\transS_{i+1}$ to pairs of edges touching the
in-frontier of $\boldS_{i+1}$ (Figs.
\ref{figure:HasseEvaluationAlone}-$i$,\ref{figure:HasseEvaluationAlone}-$ii$). 

Below we define the notion of {\em transitivity coloring} that will allow us to
perform a "sliced" transitive reduction on $DAGs$.  We say that an edge $e$ of
a slice $\boldS$ is marked by $\transS$ if $\transS(ee)=11$ and unmarked if
$\transS(ee)=00$. 

\begin{definition}[Transitivity Coloring]
\label{definition:TransitivitySeasoning} Let $\boldS=(I\cup \{v\}\cup O,E,l)$
be a unit slice. Then a {\em transitivity coloring} of $\boldS$ is a partial
function  $\transS: E^2\rightarrow \{0,1\}^2$ such that \begin{enumerate} \item
\label{item:Undefinedness}Undefinedness:  $\transS(e_1e_2) \mbox{ is not
defined if and only if } (e_1^s\!\!=\!e_2^t) \mbox{ or } (e_1^t\!\!=\!e_2^s)$
\item \label{item:Antisymmetry} Antisymmetry:  If  $\transS(e_1e_2) = ab$  then
$\transS(e_2e_1) = ba$.  \item \label{item:Marking} Marking: $\transS(ee) \in
\{00,11\}$. $e$ is unmarked if $\transS(ee)=00$ and marked if $\transS(ee)=11$.
\item \label{item:Transitivity}Transitivity: \begin{enumerate}	\item If $e_1$
and $e_2\neq e_1$ have the same source vertex, then $\transS(e_1e_2) = 00$. 
\item $\mbox{ If } e_1^s\in I \mbox{ and } e_1^t\in O
\mbox{ and } e_2^s=v \mbox{ then } \transS(e_1e_2)\in \{01 , 11\} \mbox{ and }
\\ \mbox{\hspace{3.5cm}} \transS(e_1e_2)=01 \mbox{ iff } (\exists
e,e^t\!\!=\!v)(\transS(e_1e)\in \{00, 01\})$ \end{enumerate} \item
\label{item:Relationship}Relationship between marking and transitivity:   

$$\mbox{If } e^t=v \mbox{ then } e \mbox{ is marked } \Leftrightarrow (\exists e_1, e_1^t=v)
\transS(ee_1)=01$$ 

\end{enumerate} 
\end{definition}

We observe that an isolated unit slice may be transitivity colored in many
ways. However as stated in the next lemma (Lemma
\ref{lemma:SlicedTransitiveReduction}), a unit decomposition
$\boldS_1\boldS_2\cdots\boldS_n$ of a simple DAG $H$ with a unique minimal and
a unique maximal vertices, can be coherently colored in a unique way.
Furthermore, in this unique coloring, each superfluous edge of $H$ is marked.
Later, in Lemma \ref{lemma:SlicedMultiedgeTransitiveReduction} we will provide a
generalization of Lemma \ref{lemma:SlicedTransitiveReduction} that takes DAGs
with multiple edges into consideration.

\begin{lemma}[Sliced Transitive Reduction]
\label{lemma:SlicedTransitiveReduction} Let $\boldS_1\boldS_2\cdots\boldS_n$ be
a unit decomposition of a simple DAG $H$ with a unique minimal and a unique
maximal vertices. Then \begin{enumerate} \item
\label{item1:SlicedTransitiveReduction} $\boldS_1\boldS_2\cdots\boldS_n$ has a
unique transitivity coloring $\transS_1\transS_2\cdots\transS_2$. 
\item An \label{item2:SlicedTransitiveReduction} an edge $e$ in $\boldS_i$ is
marked by $\transS_i$ if and only if $e$ is a sliced part of a superfluous edge
of $H$ (Fig. \ref{figure:HasseEvaluationAlone}-$ii$).  \end{enumerate}
\end{lemma}
\begin{proof}
Let $\transS_1\transS_2\cdots\transS_n$ be a transitivity coloring of
$\boldS_1\boldS_2\cdots \boldS_n$.  By the rule of composition of colored
slices and by conditions \ref{item:Undefinedness} to \ref{item:Relationship} of
Definition \ref{definition:TransitivitySeasoning}, the value associated by
$\transS_i$ to each two distinct edges of $\boldS_i$ are completely determined
by the values associated by $\transS_{i-1}$ to edges touching the out-frontier
of $\boldS_{i-1}$. Furthermore, since $H$ has a unique minimal vertex,
$\transS_1$ associates the value $00$ to each two distinct edges of $\boldS_1$.
Thus the values associated by each $\transS_i$ to distinct edges of $\boldS_i$
are unique. It remains to show that the marking is unique. 

Let $e$ be a superfluous edge of $H$, and $e_1e_2..e_k$ be a path from $e^s$ to
$e^t$, then the transitivity conditions in Definition
\ref{definition:TransitivitySeasoning}.\ref{item:Transitivity} assure that for
any sliced part $e'$ of $e$ and any sliced part $e_i'$ of $e_i$ lying in the
same slice $\boldS_j$, $\transS_j(e',e_i')=00$ if $i=1$ and
$\transS_j(e',e_i')=01$ for $2\leq i \leq k$ (Fig.
\ref{figure:HasseEvaluationAlone}).  Let $S_j$ be the slice that contains the
target vertex of $e$, and let $e'$ and $e_k'$ be respectively the sliced parts
of $e$ and $e_k$ lying in $S_j$. Then $\transS(e'e'_k)=01$ and thus, by
condition  \ref{item:Relationship} of Definition
\ref{definition:TransitivitySeasoning}, $e'$ is marked, implying that any
sliced part of $e$ lying in previous slices must be marked as well.  Now
suppose that $e_1$ and $e_2$ have the same target vertex, and that $e_1$ is not
superfluous. Then for any sliced part $e_1'$ of $e_1$ and any sliced part
$e_2'$ of $e_2$ lying in the same slice $S_j$, we must have
$\transS_j(e_1',e_2')=10$ if $e_2$ is superfluous and $\transS_j(e_1',e_2')=11$
if $e_2$ is not superfluous. Thus by condition \ref{item:Relationship} of
Definition \ref{definition:TransitivitySeasoning}, no sliced part of $e_1$ can
be marked.  We observe that $\transS_j(e_1',e_2')\neq 00$ since otherwise $e_1$
and $e_2$ would have the same source and thus form a multiple edge. $\square$
\end{proof}

Next, in Theorem \ref{theorem:TransitiveReductionSliceGraphsA} we deal with the
transitive reduction of slice graphs that generate only DAGs without multiple
edges, which we call {\em simple slice graphs}. Lemma \ref{lemma:SlicedTransitiveReduction} is of special
importance for its proof. The transitive reduction of general slice graphs will
be addressed in Theorem \ref{theorem:TransitiveReductionSliceGraphsB}. 

\begin{theorem}[Transitive Reduction of Simple Slice Graphs]
\label{theorem:TransitiveReductionSliceGraphsA} Let $\slicegraph=(\mathcal{V},
\mathcal{E},\mathcal{S})$ be a slice graph such that $\lang_G(\slicegraph)$ has
only simple DAGs.  Then there exists a Hasse diagram generator
$\hasseGenerator$ such that $\lang_{PO}(\slicegraph) =
\lang_{PO}(\hasseGenerator)$.  \end{theorem} \begin{proof} As a first step we
construct an intermediary slice graph $\slicegraph'$ as follows: we expand each
vertex $v$ in $\mathcal{V}$ with a set of vertices $\{v_{\transS}\}$ where
$\transS$ ranges over all transitivity colorings of $\mathcal{S}(v)$. Each
vertex in $\{v_{\transS}\}$ is labeled with $\mathcal{S}(v)$. We add an edge
from $v_{\transS}$ to $v'_{\transS'}$ in $\slicegraph'$ if and only if $v$ is
connected to $v'$ in $\slicegraph$ and if the values associated by $\transS$ to
the edges touching the out-frontier of $\mathcal{S}(v)$ agree with the values
associated by $\transS'$ to the edges touching the in-frontier of
$\mathcal{S}(v')$. Finally we delete vertices that cannot be reached from an
initial vertex, or that cannot reach a final vertex.  We note that
$\transS_1\transS_2\cdots\transS_n$ is a transitivity coloring of the label
$\mathcal{S}(v^1) \mathcal{S}(v^2) \cdots \mathcal{S}(v^n)$  of a walk from a
initial vertex $v_1$ to a final vertex $v_2$ in $\slicegraph$, if and only if
$\mathcal{S}(v^1) \mathcal{S}(v^2)\cdots\mathcal{S}(v^n)$ also labels the walk
$v^1_{\transS_1}v^2_{\transS_2}\cdots v^n_{\transS_n}$ in the new slice graph
$\slicegraph'$. By Lemma
\ref{lemma:SlicedTransitiveReduction}.\ref{item1:SlicedTransitiveReduction} a
coloring exists for each such a walk and thus
$\lang_G(\slicegraph)=\lang_G(\slicegraph')$.  In order to get the Hasse
diagram generator $\hasseGenerator$ with the same partial order language as
$\slicegraph$, we relabel each vertex $v_{\transS}$ with a version of
$\mathcal{S}(v)$ in which the edges which are marked by $\transS$ are deleted.
By Lemma
\ref{lemma:SlicedTransitiveReduction}.\ref{item2:SlicedTransitiveReduction} a
DAG is in $\lang_G(\hasseGenerator)$ if and only if it is  the transitive
reduction of a $DAG$ in $\slicegraph$, and thus
$\lang_{PO}(\hasseGenerator)=\lang_{PO}(\slicegraph)$.  
$\square$
\end{proof}

In general the graph language of a slice graph may contain DAGs with multiple
edges. Below we extend Definition \ref{definition:TransitivitySeasoning} to
deal with these DAGs. Given a unit decomposition
$\boldS_1\boldS_2\cdots\boldS_n$ of a DAG $H$, we partition each frontier of
$\boldS_i$ into numbered cells in such a way that two edges touch the same cell
of a frontier if and only if they are the sliced parts of edges with the same
source and target in $H$. 

\begin{definition}[Multi-edge Transitivity Coloring]
\label{definition:Multiedge} A multi-edge transitivity coloring of a unit slice
$\boldS=(I\cup \{v\}\cup O,E,l)$ is a triple
$(\transS,\partition^{in},\partition^{out})$ where $\transS$ is a transitivity
coloring of $\boldS$, $\partition^{in}$ is a numbered partition of the
in-frontier vertices $\boldS$ and $\partition^{out}$ a numbered partition of
the out-frontier vertices of $\boldS$, such that: \begin{enumerate} \item
\label{item0:Multiedge} If two edges touch the same cell in one of the
partitions then either both are connected to $v$ or both touch the same cell in
the other partition.  \item \label{item1:Multiedge} For any edge $e$ and any
distinct edges $e_1$ and $e_2$ touching the same cell of $\partition^{in}$ or
the same cell of $\partition^{out}$, $\transS(ee_1)=\transS(ee_2)$.  \item
\label{item2:Multiedge} In each cell of $\partition^{in}$ and in each cell of
$\partition^{out}$ either all edges are marked or all edges are unmarked.
\item \label{item3:Multiedge} If $v$ is the target of two edges $e_1, e_2$ and
$\transS(e_1e_2)=00$ then $e_1$ and $e_2$ belong to the same cell of
$\partition^{in}$.  \end{enumerate} \end{definition}

A sequence
$(\transS_1,\partition_1^{in},\partition_1^{out})(\transS_2,\partition_2^{in},\partition_2^{out})\cdots(\transS_n,\partition_n^{in},\partition_n^{out})$
is a multi-edge transitivity coloring of a unit decomposition
$\boldS_1\boldS_2\cdots\boldS_n$ of a DAG $H$, if $\transS_1\transS_2\cdots
\transS_n$ is a transitivity coloring of $\boldS_1\boldS_2\cdots\boldS_n$ and
for each $i$, two edges $e_1,e_2$ in $\boldS_i$ touch the same cell of
$\partition^{out}_i$ if the corresponding edges to which they are glued in
$\boldS_{i+1}$ touch the same cell of $\partition^{in}_{i+1}$.

We note in special that condition \ref{item3:Multiedge} of Definition
\ref{definition:Multiedge} will guarantee that in a coherent multi-edge
transitivity coloring of a unit decomposition of a DAG, all sliced parts of
multiple edges with the same source and target in the DAG, will touch the same
cells in both of the partitions. This happens because in each transitivity
colored slice of the decomposition, two distinct edges with the same target are
colored with the value $00$ if and only if they are the sliced parts of
multiple edges in the original DAG. From our discussion, we may state an
adapted version of Lemma \ref{theorem:TransitiveReductionSliceGraphsA} that
takes multiple edges into account (Lemma
\ref{lemma:SlicedMultiedgeTransitiveReduction}). In this case there may be more
than one valid coloring, but they differ only in the way the vertices of the
frontier of each slice are partitioned.  

\begin{lemma}[Sliced Multi-edge Transitive Reduction]
\label{lemma:SlicedMultiedgeTransitiveReduction} Let
$\boldS_1\boldS_2\cdots\boldS_n$ be a unit decomposition of a DAG $H$ with a
unique minimal and maximal vertices. Then \begin{enumerate} \item
\label{item1:SlicedTransitiveReduction} $\boldS_1\boldS_2\cdots\boldS_n$ has at
least one multi-edge transitivity coloring
$$(\transS_1,\partition_1^{in},\partition_2^{out})(\transS_2,\partition_2^{in},\partition_2^{out})
\cdots(\transS_n,\partition_n^{in},\partition_n^{out}).$$ Furthermore,  \item
\label{item2:SlicedTransitiveReduction} an edge $e$ in $\boldS_i$ is marked by
$\transS_i$ if and only if $e$ is a sliced part of a superfluous edge of $H$
(Fig. \ref{figure:HasseEvaluationAlone}-$ii$).  \item Two edges $e_1,e_2$ of
$\boldS_i$ touch the same cell of $\partition_i^{in}$ or the same cell of
$\partition_i^{out}$ if and only if they are sliced parts of edges $e'_1,e'_2$
of $H$ with same source and tail vertices.  \end{enumerate} \end{lemma}

As a consequence our transitive reduction algorithm described in Theorem
\ref{theorem:TransitiveReductionSliceGraphsA} may be adapted to work with
general slice graphs, and not only with those that generate simple DAGs. 

\begin{theorem}[Transitive Reduction of General Slice Graphs]
\label{theorem:TransitiveReductionSliceGraphsB} Let $\slicegraph=(\mathcal{V},
\mathcal{E},\mathcal{S})$ be a slice graph.  Then there exists a Hasse diagram
generator $\hasseGenerator$ such that $\lang_{PO}(\slicegraph) =
\lang_{PO}(\hasseGenerator)$.  \end{theorem} \begin{proof} The proof proceeds
as in the proof of Theorem \ref{theorem:TransitiveReductionSliceGraphsA}. To
avoid a cumbersome notation we write $\partition$ for the pair
$(\partition^{in},\partition^{out})$. In the construction of the intermediary
slice graph $\slicegraph'$, we expand each vertex $v$ with a set of vertices
$\{v_{\transS,\partition}\}$, and label each of them with $\mathcal{S}(v)$. We
connect $v_{\transS\partition}$ to $v'_{\transS'\partition'}$ if and only if
$v$ is connected to $v'$ in $\slicegraph$ and both and the cells of
$\partition$ and the and the values of $\transS$ on the out-frontier of
$\mathcal{S}(v)$ agree with the cells of $\partition'$ and the values of
$\transS'$ on the in-frontier of $\mathcal{S}(v')$. The only point that differs
in the proof is the relabeling of the vertices of $\slicegraph'$ in order to
transform it into a Hasse diagram generator. Namely, each vertex
$v_{\transS,\partition}$ of $\slicegraph'$ is relabeled with a version of
$\mathcal{S}(v)$ in which for each $i$, the $i$-th cell of the partition
$\partition^{in}$ ($\partition^{out}$) is collapsed into a single in-frontier
(out-frontier) vertex labeled by $i$, all edges touching the same cell of a
partition are collapsed into a unique edge (Fig.
\ref{figure:TransitiveReductionSliceGraphs}.$i$), and all the marked edges are
deleted. By Lemma \ref{lemma:SlicedMultiedgeTransitiveReduction} the DAGs
generated by $\hasseGenerator$ are the transitive reduced counterparts of the
DAGs generated by $\slicegraph$ and consequently,
$\lang_{PO}(\hasseGenerator)=\lang_{PO}(\slicegraph)$.  
$\square$
\end{proof}

We end this section by giving a simple upper bound on the complexity of the
transitive reduction of slice graphs: 

\begin{corollary} \label{corollary:Complexity} Let $\slicegraph$ be a slice
graph with $n$ vertices and let $s$ be the size of the greatest frontier of a
slice labeling a vertex of $\slicegraph$. Then the Hasse diagram generator
constructed in in Theorem \ref{theorem:TransitiveReductionSliceGraphsB} has
$n\cdot 2^{O(s^2)}$ vertices. In particular, the transitive reduction algorithm
runs in polynomial time for $s=O(\sqrt{\log n})$.  \end{corollary}
\begin{proof}
Let $v$ be a vertex of $\slicegraph$ which is labeled with a slice $\boldS$ of 
width $s$. Then there are at most $2^{O(s^2)}$ transitivity colorings of $\boldS$,
since each two edges touching the same frontier of $\boldS$ can be colored in at 
most a constant number of ways. Furthermore, there are at most $2^{O(s\log s)}$ 
possible ways of partitioning a set of size $s$, and thus of partitioning each 
frontier of $\boldS$. This implies that the number of possible multi-edge transitive 
colorings of $\boldS$ is still bounded by $2^{O(s^2)}$. Since $\slicegraph$ 
has $n$ vertices, the bound of $n\cdot 2^{O(s^2)}$ follows. $\square$
\end{proof}

%%%%%%%%%%%%%%%%%%%%%%%%%%%%%%%%%%%%%%%%%%%%%%%%%%%%%%%%%%%%%%%%%%%%%%%%%%%%%%%%
%%%%%%%%%%%%%%%%%%%%%%%%%%%%%%%%%%%%%%%%%%%%%%%%%%%%%%%%%%%%%%%%%%%%%%%%%%%%%%%%

\section{Saturated and Weakly Saturated Slice Languages} \label{section:Saturation}

In this section we introduce {\em weakly saturated slice languages} and show that they allow 
us to smoothly generalize regular string languages to the partial order setting. In particular, 
the class of partial order languages which are definable through weakly saturated slice 
languages is closed under union intersection and even under a suitable notion of complementation, which we call 
{\em globally bounded complementation}. Furthermore, both inclusion and emptiness
of intersection are decidable for this class. 

Weakly saturated slice languages generalize both recognizable Mazurkiewicz trace languages
\cite{Mazurkiewicz1986} and linearization-regular message sequence chart languages
\cite{HenriksenMukundKumarSohoniThiagarajan2005}.  It turns out that this
generalization is strict. As showed by us in \cite{deOliveiraOliveira2010},
regular slice languages are expressive enough to represent the partial order 
behavior of any bounded $p/t$-net. Furthermore, as we will prove in Section \ref{section:PetriNets},
the Hasse diagram generators associated to $p/t$-nets in \cite{deOliveiraOliveira2010} are saturated. 
In contrast with this result, we note that the partial order behavior of bounded $p/t$-nets cannot be represented by 
Mazurkiewicz traces, which for example rule out auto-concurrency, neither 
by MSC languages. Indeed these formalisms 
are not able to capture even the partial order behavior of $1$-bounded $p/t$-nets. On the other hand, as we 
will show in Section \ref{section:Reductions}, both Mazurkiewicz traces 
and MSC languages can be reinterpreted in terms of non-transitive 
reduced slice languages, and through an application of our transitive 
reduction algorithm, they can be indeed mapped to Hasse diagram generators representing the 
same set of partial orders.

\begin{figure}[t] \centering
\includegraphics[scale=0.33]{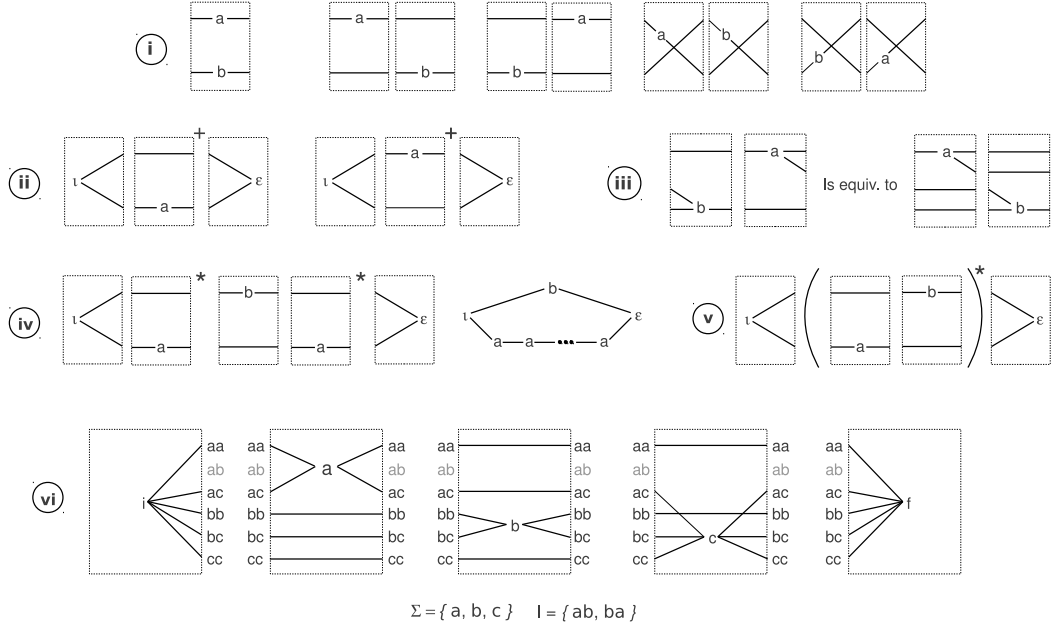}
\caption{ $i$) A slice and its set of unit decompositions.
\label{figure:slicingSet} $ii$)Two regular expressions over slices generating
the same graph and partial order languages, but distinct slice languages.
\label{figure:notConverseImplication}$iii$)   \label{figure:mazurkiewicz}
Equivalence of slice strings.  $iv$)\label{figure:WeaklySaturated} A slice
expression over slices generating a weakly saturated slice language and an
intuitive depiction of its graph language. $v$) An example of slice language
that cannot be saturated. \label{figure:NotSaturable} $vi$) Mapping an
independence alphabet to a slice alphabet. \label{figure:TracesToSlices} }
\label{figure:sliceStrings} \end{figure}

The following chain of implications relating the graph and partial order
languages represented by two slice languages $\lang$ and $\lang'$ is a direct
consequence of Equation (\ref{equation:graphLanguage}):

\begin{equation} \label{equation:ChainImplications} \lang \subseteq
\lang'\Rightarrow  \lang_{G} \subseteq \lang'_{G} \Rightarrow \lang_{PO}
\subseteq \lang'_{PO} \end{equation}

However there are simple examples of slice languages $\lang$ and $\lang'$ for
which $\lang_G\subseteq \lang'_G$ and $\lang\nsubseteq \lang'$ (Fig.
\ref{figure:notConverseImplication}.$ii$) or for which $\lang_{PO}\subseteq
\lang_{PO}'$ and $\lang_G\nsubseteq \lang_G'$ (Fig.
\ref{figure:TransitiveReductionSliceGraphs}). If $\lang$ and $\lang'$ are
regular slice languages, then the inclusion $\lang\subseteq \lang'$ can be
decided by standard finite automata techniques. However, even if $\lang$ and
$\lang'$ are regular slice languages it is undecidable whether
$\lang_{G}\subseteq \lang'_{G}$ ($\lang_{PO}\subseteq \lang'_{PO}$) as well as
whether  $\lang_{G} \cap \lang'_{G} =\emptyset$ ($\lang_{PO}\cap
\lang_{PO}'=\emptyset$) \cite{deOliveiraOliveira2010}. Indeed since slice
languages strictly generalize trace languages (Section
\ref{section:Reductions}), these undecidability results may be regarded as
an inheritance from analogous results in trace theory
\cite{Ibarra1978,AalbersbergHoogeboom1989}. Fortunately, these and other
related problems become decidable for the weakly saturated saturated slice languages, 
which we define below (Definition \ref{definition:WeaklySaturatedSliceGraphs}). Before, 
recall that a topological ordering of a DAG $H=(V,E,l)$ is an ordering $v_1v_2\cdots v_n$
of its vertices such that $i< j$ whenever $v_i< v_j$ in the partial ordering
induced by $H$. 

\begin{definition}[Weakly Saturated Slice Graphs]
\label{definition:WeaklySaturatedSliceGraphs}
 We say that a slice language $\lang$ is weakly saturated if
for every DAG $H\in \lang_G$ and every topological ordering $v_1v_2\cdots v_n$
of $H$, $\lang$ has a unit decomposition $\boldS_1\boldS_2\cdots\boldS_n$ of
$H$ in which $v_i$ is the center vertex of $\boldS_i$, for each $1\leq i\leq n$.
A slice graph is weakly saturated if it generates a weakly saturated slice
language. 
\end{definition}

An important property of our transitive reduction algorithm (Theorem \ref{theorem:TransitiveReductionSliceGraphsB})
is that it preserves weak saturation:

\begin{proposition}[Transitive Reduction Preserves Weak Saturation]
\label{proposition:PreservesSaturation}
Let $\slicegraph$ be a weakly saturated slice graph over the alphabet $\slicealphabet^c$ 
and $\hasseGenerator$ be the transitive reduced version of $\slicegraph$ after the application 
of Theorem \ref{theorem:TransitiveReductionSliceGraphsB}. Then $\hasseGenerator$
is weakly saturated. 
\end{proposition}
\begin{proof}
The proof follows from the fact that an ordering $v_1,v_2,...,v_n$ of the vertices
of a $DAG$ $G$ on $n$ vertices is a topological ordering of $G$ if and only if it is
also a topological ordering of the Hasse diagram of $G$. $\square$
\end{proof}

Even though {\em weak saturation} is the concept that is meant to be used 
in practice, in proofs it will be more convenient to deal with the notion of 
{\em saturation}, which we define below (Definition \ref{definition:SaturatedSliceLanguage}). 
In the end of this section (Theorem \ref{theorem:WeakSaturation}) we will show that each weakly saturated slice graph can 
be efficiently transformed into a saturated one generating the same partial 
order language, and thus all decidability results that are valid for the latter class
of slice graphs are also valid for weakly saturated slice graphs. In Figure \ref{figure:WeaklySaturated}.$iv$ we depict a regular
expression over slices generating a weakly saturated (but not saturated) slice
language. 

\begin{definition}[Saturated Slice Languages]
\label{definition:SaturatedSliceLanguage} Let $ud(\boldS)$ ($ud(H)$) be the
set of all unit decompositions of a slice $\boldS$ (a DAG $H$) (Fig.
\ref{figure:slicingSet}.$i$).  We say that a slice language $\lang$ is {\em
saturated } if $u(H)\subseteq \lang$ whenever $H\in \lang_G$.  We say that a
slice graph $\slicegraph$ is saturated if it generates a saturated slice
language.  \end{definition}

It turns out that the saturation of Definition \ref{definition:SaturatedSliceLanguage} 
can be restated in terms of the closure of a slice language, 
under a notion of commutation defined on its slice alphabet. Suppose 
that any unit decomposition of a $DAG$ in the graph language represented by a 
slice language $\lang$ has slice width at most $c$. Let $\slicealphabet^c$ be 
the set of all unit slices of slice width at most $c$\;\footnote{More precisely 
$\slicealphabet^c(T)$ for some set of events $T$.} (Section \ref{section:Slices}). 
We say that two unit slices $\boldS$ and $\boldS'$ in $\slicealphabet^c$ are independent of each other if 
there is no edge joining the center vertex $v$ of $\boldS$ to the center vertex $v'$ 
of $\boldS'$ in the slice $S\circ S'$ (Fig. \ref{figure:mazurkiewicz}.$iii$).  Let $u$ and $w$
be strings over $\slicealphabet^c$. We say that the slice string
$u\boldS_1\boldS_2w$ is similar to $u\boldS_1'\boldS_2'w$ ($u\boldS_1\boldS_2w
\simeq u\boldS_1'\boldS_2'w$) if $\boldS_1\circ \boldS_2=\boldS'_1\circ
\boldS'_2$. The reflexive and transitive closure $\simeq^*$ of $\simeq$ is an
equivalence class over slice strings. If the composition of the slices in a
slice string $\boldS_1\boldS_2\cdots\boldS_n$ gives rise to a DAG $H$, then the
class of equivalence in which $\boldS_1\boldS_2\cdots\boldS_n$ lies is equal to
the set of unit decompositions of $H$, i.e. $ud(H)$.  We observe that not necessarily every word in the
free monoid generated by $\slicealphabet^c$ corresponds to a valid graph. This 
however is not a problem when it comes to slice languages generated by slice graphs, 
and our equivalence relation on slices, gives us a way to test whether a regular slice 
language $\lang$ is saturated. As we show in the next theorem, it suffices to determine 
whether the minimal finite automaton generating $\lang$ is "diamond" closed. In Section 
\ref{subsection:MazurkiewiczTraces} we will compare our notion of independence, 
with the notion of independence used in Mazurkiewicz trace theory. 

\begin{theorem} 
\label{theorem:DecidingSaturation}
Let $\slicegraph$ be a slice graph over a slice alphabet
$\slicealphabet^c$. Then we may effectively determine whether the slice language
generated by $\slicegraph$ is saturated.  
\end{theorem} 
\begin{proof} Let $c$ be the size of the largest slice labeling a vertex of $\slicegraph$. Thus 
the slice languages generated by $\slicegraph$ is a subset of $\slicealphabet^c$. In order
to verify whether a slice graph $\slicegraph$ generates a saturated slice
language, it is enough to test the following condition: If a slice word
$w\boldS_1\boldS_2 u$ is generated by $\slicegraph$ then every word
$w\boldS_1'\boldS_2'u$ satisfying $\boldS_1'\circ\boldS_2'=
\boldS_1\circ\boldS_2$ is generated by $\slicegraph$ as well.  Let
$\sliceautomaton$ be the minimal deterministic finite automaton over
$\slicealphabet$ that generates the same slice language as $\slicegraph$. Since
the automaton is minimal and deterministic, any string
$\boldS_1\boldS_2\cdots\boldS_k\in \lang(\sliceautomaton)$ corresponds to a
unique computational path of $\sliceautomaton$. In particular this implies that
to verify our condition, we just need to determine whether $\sliceautomaton$ is 
"diamond" closed. In other words we need to test whether for each pair of transition rules $q\boldS_1 r$ and $r\boldS_2 q'$ of the
automaton and each unit decomposition $\boldS_1'\boldS_2'$ of
$\boldS_1\circ\boldS_2$, the automaton has a state $r'$ and transitions
$q\boldS_1'r'$ and $r'\boldS_2'q'$. Clearly this condition can be effectively
verified efficiently, since $\boldS_1\circ \boldS_2$ can have at most a
polynomial (on the size of $\boldS_1\circ \boldS_2$) number of unit
decompositions. $\square$ \end{proof}

\begin{figure}[h]
\centering 
\includegraphics[scale=0.35]{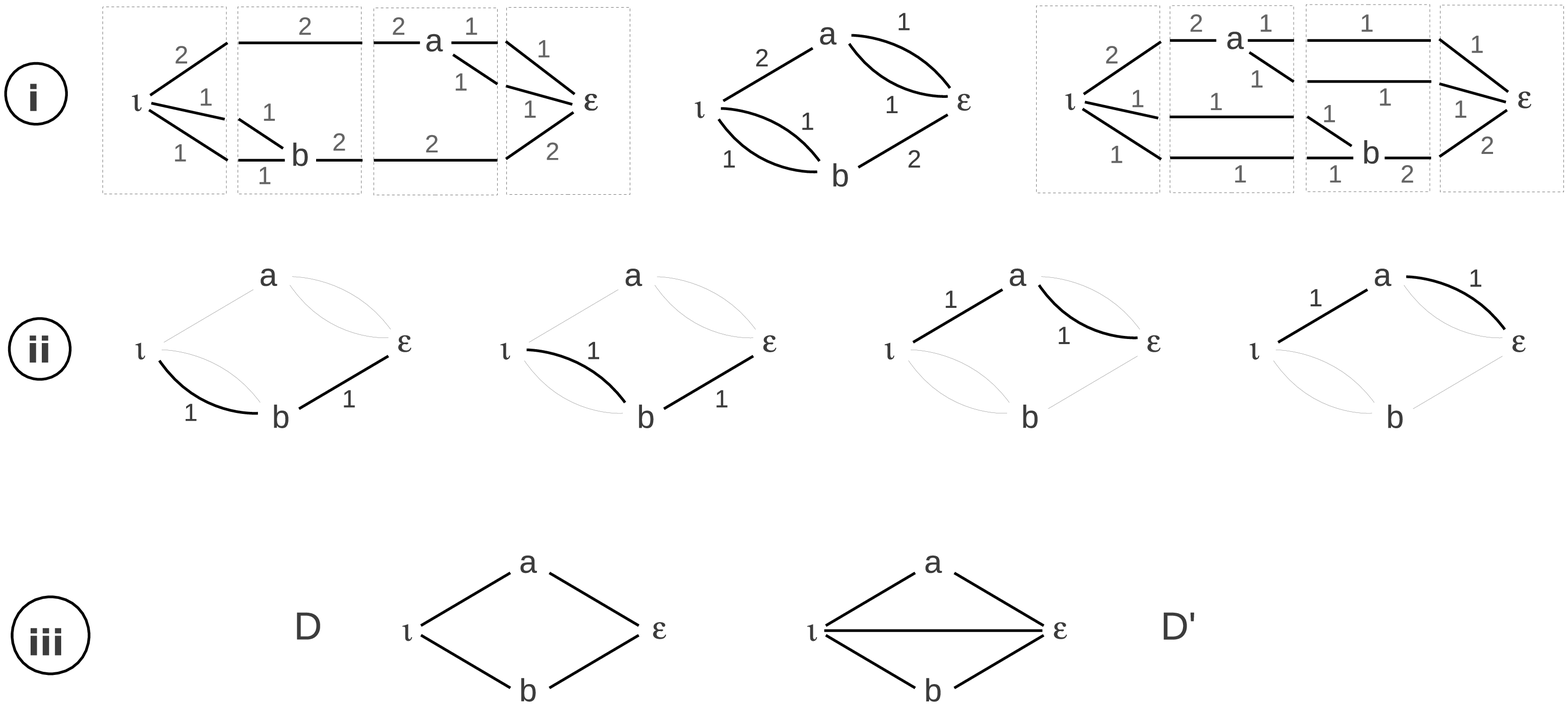}
\caption{$i)$ A DAG $G$ whose edges are colored with a $4$-flow $f$, and two unit decompositions of $G$ colored with sliced versions of $f$. 
Suppose that $\lang$ is a slice language and that $G\notin \lang_G$. Then the unit decomposition to the left belongs to 
$(\slicealphabet^c)^*\backslash \lang$ for $c=3$, but not the unit decomposition to the right, which has slice width $4$. $ii)$ A $c$-flow 
$f$ can be regarded as the sum of $c$ unit flows (In this case $c=4$). $iii)$ A diamond DAG $D$ and a diamond with an additional edge $D'$.
Let $\lang=ud(D)$ and $\lang'=ud(D')$ then both $\lang$ and $\lang'$ are saturated and $\lang_{PO}(D) = \lang_{PO}(D') \neq \emptyset$ but $\lang \cap \lang'=\emptyset$
and $\lang_G\cap \lang_G'=\emptyset$. } 

\label{figure:langNotSaturated}
\end{figure}

The class of graphs which can be represented by saturated slice languages
is closed under union and intersection and has decidable inclusion and emptiness of 
intersection. Indeed these facts follow from the following equation, which is 
valid for saturated slice languages:  
\begin{equation}
\label{equation:SaturatedUnion}
	\lang = \bigcup_{H\in \lang_G} ud(H)
\end{equation}

where $ud(H)$ denotes the set of all unit decompositions of $H$. 
The complement of graph languages representable by saturated slice graphs is more subtle, 
and {\bf does not} follow directly from equation \ref{equation:ComplementationSliceLanguages} nor from the 
commutation operation defined on $\slicealphabet^c$. We say that a DAG $G$ has 
{\em global slice width} $c$, if every unit decomposition of $G$ has slice width at most $c$. 
Now suppose that a language $\lang_G$ of graphs can be represented by a regular and saturated slice language 
$\lang \subseteq (\slicealphabet^c)^*$. Then we define the complement of $\lang_G$ to be

\begin{equation}
\label{equation:ComplementationSliceLanguages}
\overline{\lang_G} = \lang^c_G \backslash \lang_G
\end{equation}

where $\lang^c_G$ is the language consisting of all DAGs of global slice width at most $c$.
The caveat is that the fact that $\lang^c_G$ can be represented by a saturated regular slice language
is not evident at all. Intuitively one could expect that the closure of $\lang$ under commutation 
would imply that the complement of $\lang_G$ could be represented at a syntactic level by intersecting $(\slicealphabet^c)^*\backslash \lang$
with the set of all legal\footnote{By legal we mean slice strings which can be composed to form a DAG.} 
slice strings over $\slicealphabet^c$. As illustrated in figure \ref{figure:langNotSaturated} this intuition 
is misleading, and indeed $(\slicealphabet^c)^*\backslash \lang$ may generate graphs of global slice width greater 
than $c$. The construction of a saturated slice graph $\slicegraph^c$ generating $\lang_G^c$ will be carried in the next subsection 
(Subsection \ref{subsection:UniversalSliceGraph}), and will follow from a characterization of 
graphs of global slice width in terms of flows. 

A similar nuance will appear when defining a suitable notion for the complementation of the 
partial order language $\lang_{PO}$ represented by a saturated slice language $\lang$. We 
define the $c$-globally bounded complementation of $\lang_{PO}$ to be the partial order 
language

\begin{equation}
\label{equation:ComplementationPartialOrderLanguages}
\overline{\lang}_{PO} = \lang^c_{PO} \backslash \lang_{PO}
\end{equation}

where $\lang_{PO}^{c}$ is the partial order language induced by $\lang_{G}^c$. 
As we will show in the next subsection, $\lang_{PO}^c$ can be represented by 
a saturated Hasse diagram generator $\hasseGenerator^c$. Four ingredients will be 
essential for the construction of $\hasseGenerator^c$. First, the fact 
mentioned above that $\lang_G^c$ can be represented by a saturated slice graph over 
$\slicealphabet^c$. Second our transitive reduction 
algorithm (Theorem \ref{theorem:TransitiveReductionSliceGraphsB}) which will be applied to $\slicegraph^c$. 
Third, the fact that transitive reduction preserves weak saturation (Proposition \ref{proposition:PreservesSaturation}) and 
finally the fact that weakly saturated slice languages can be transformed into saturated slice languages (Theorem \ref{theorem:WeakSaturation}).

\subsection{Globally Bounded Slice Graph and Globally Bounded Hasse Diagram Generator}
\label{subsection:UniversalSliceGraph}

In order to construct the slice graph $\slicegraph^c$ (Lemma \ref{lemma:ConstructionGlobalSliceWidth}) representing $\lang_{PO}^c$, 
we will need to introduce a "sliced characterization" of graphs of global slice width $c$. With this goal in mind, we define 
the notion of $c$-flow coloring for unit slices:

\begin{definition}[$c$\;-flow coloring]
\label{definition:CFlowColoring}
Let $\boldS=(\{v\},E,l)$ be a unit slice. We say that a function $f:E\rightarrow \N$ is a $c$-flow coloring of $\boldS$ 
if $f:E\rightarrow \N$ satisfies the following conditions: 
\begin{enumerate}
	\item \label{CFlowColoring:Item:Positivity} Positivity: \label{item:GreaterThanOne} For any edge $e\in E$, $f(e)\geq 1$,
	\item \label{CFlowColoring:Item:Conservativity} Vertex Conservativity: \label{item:Conservative}
		\begin{itemize} 
			\item If both frontiers of $\boldS$ are non-empty, then $\sum_{e_1^s=v} f(e) = \sum_{e_2^t=v}f(e)$
		\end{itemize}
	\item \label{CFlowColoring:Item:Propagation} Frontier Conservativity: 
		\begin{itemize}
			\item If the in-frontier of $\boldS$ is non-empty then $\sum_e f(e)=c$ where
		$e$ ranges over the edges touching the in-frontier of $\boldS$. 
			\item If the out-frontier of $\boldS$ is non-empty 
		then $\sum_{e'}f(e')=c$ where $e'$ ranges over the edges touching the out-frontier of $\boldS$.
		\end{itemize}
\end{enumerate}
\end{definition}

A $c$-flow coloring of a unit decomposition $\boldS_1\boldS_2...\boldS_n$ of a DAG $G$, is a sequence of functions $f_1f_2...f_2$ such that 
each $f_i$ is a $c$-flow coloring of $\boldS_i$, and such that the values associated to edges touching the out-frontier of $\boldS_i$ 
agree with the values associated by $f_{i+1}$ to edges touching the in-frontier of $\boldS_{i+1}$ (Fig \ref{figure:langNotSaturated}.$i$).
In Lemma \ref{lemma:GlobalSliceWidth} below we assume that the DAGs have a unique minimal and a unique maximal vertex. This assumption 
is not at all essential and is made only for the sake of avoiding the consideration of several special cases. 

\begin{lemma}[$c$\;-Flows and Global Slice Width]
\label{lemma:GlobalSliceWidth}
Let $G=(V,E,l)$ be a DAG with a unique minimal vertex $v_{\iota}$ and a unique maximal vertex $v_{\varepsilon}$. Then $G$ has global slice width 
at most $c$ if and only if there exists a function $f:E\rightarrow \N$ satisfying the following conditions: 
\begin{enumerate}
	\item Positivity: \label{GlobalSliceWidth:Item:Positivity} For any edge $e$ of $G$, $f(e)\geq 1$,
	\item Vertex Conservativity: \label{GlobalSliceWidth:Item:Conservative} For every $v\in V$, $\sum_{e_1^s=v} f(e) = \sum_{e_2^t=v}f(e)$
	\item Initialization and Finalization: \label{GlobalSliceWidth:Item:InitialConditions} $\sum_{e^s=v_{\iota}} f(e) = c = \sum_{e^t=v_{\varepsilon}} f(e)$
\end{enumerate}
\end{lemma}
\begin{proof}
Let $G$ be a DAG and $f:E\rightarrow \N$ be a function satisfying conditions \ref{GlobalSliceWidth:Item:Positivity} to \ref{GlobalSliceWidth:Item:InitialConditions}. 
To each slice decomposition $\boldS_1\boldS_2...\boldS_n$ of $G$ where $\boldS_i=(V_i,E_i,l_i)$, we may associate a $c$-flow coloring 
$f_1f_2...f_n$ (with  $f_i:E_i\rightarrow \N$) by setting $f_i(e)=f(e')$ whenever $e\in E_i$ is a sliced part of the edge $e\in E$. Clearly
each $f_i$ satisfies conditions \ref{CFlowColoring:Item:Positivity} and \ref{CFlowColoring:Item:Conservativity} of Definition \ref{definition:CFlowColoring}. 
Condition \ref{CFlowColoring:Item:Propagation} follows by induction on $i$. It holds for $i=1$ by the Initialization condition of the present lemma, 
and holds for $i>1$ by noticing that the sum of values associated to edges the in-frontier of a slice $\boldS_i$ must be equal 
to the sum of values associated to the edges in the out-frontier of $\boldS_{i-1}$ (Fig \ref{figure:langNotSaturated}). Now since 
each $f_i$ is a $c$-coloring of $\boldS_i$, by conditions \ref{CFlowColoring:Item:Positivity} and \ref{CFlowColoring:Item:Propagation}, 
each frontier of $\boldS_i$ has at most $c$ edges, and thus $G$ has global slice width at most $c$.  

Now Suppose that $G$ has global slice width $c$. Since $G$ has a unique minimal vertex $v_{\iota}$ and a unique maximal 
vertex $v_{\varepsilon}$, we have that $G$ can be cast as the union of $c$ paths (not necessarily disjoint paths) from $v_{\iota}$ to $v_{\varepsilon}$. 
Let $G=\cup_{w} w$ be the union of these $c$ paths from $v_{\iota}$ to $v_{\varepsilon}$ (Fig. \ref{figure:langNotSaturated}.$ii$). 
For each such a path $w= e_1 e_2...e_{n-1}e_n \in G$ consider the function $f_w:E\rightarrow \N$ that associates the value $1$ to each 
edge in $w$ and the value $0$ to each edge of $G$ which is not in $w$. We claim that the function $f=\sum_{w\in W} f_w$ is a $c$-flow of $G$:
It clearly satisfies Condition \ref{GlobalSliceWidth:Item:Positivity}, since each edge of $G$ belongs to at least one path. Condition 
\ref{GlobalSliceWidth:Item:Conservative} follows from the fact that for each intermediary vertex of the path both the edge which arrives 
to $v$ and the edge that departs from $v$ receive the value $1$. Condition \ref{GlobalSliceWidth:Item:InitialConditions} follows from the 
fact that every considered path starts at $v_{\iota}$ and finishes at $v_{\varepsilon}$. $\square$

\end{proof}

%%%%%%%%%%%%%%%%%%%%%%%%%%%%%%%%%%%%%%%%%%%%%%%%%%%%%%%%%%%5

\begin{corollary}[Sliced Characterization of Global Slice Width]
\label{corollary:SlicedCharacterizationGlobalSliceWidth}
Let $G=(V,E,l)$ be a DAG with a unique minimal vertex and a unique maximal vertex. Then 
$G$ has global slice width $c$ if and only if each slice decomposition $\boldS_1\boldS_2...\boldS_n$
of $G$ has a $c$-flow coloring $f_1f_2...f_n$. 
\end{corollary}
\begin{proof}
By Lemma \ref{lemma:GlobalSliceWidth}, $G$ has global slice width $c$ if and only if $G$ admits a $c$-flow $f$. Let $\boldS_1\boldS_2...\boldS_n$ be a unit decomposition 
of $G$. Then as shown in the proof of Lemma \ref{lemma:GlobalSliceWidth}, if we let $f_i:E_i\rightarrow \N$ and set $f_i(e)=f(e')$ whenever 
$e\in E_i$ is a sliced part of $e\in E$, then $f_1f_2...f_n$ is a $c$-flow coloring of $\boldS_1\boldS_2...\boldS_n$ (Fig \ref{figure:langNotSaturated}).
\end{proof}

\begin{lemma}[Globally Bounded Slice Graph]
\label{lemma:ConstructionGlobalSliceWidth}
For each $c\in \N$ with $c\geq 1$ there is a saturated slice graph $\slicegraph^c$ on $2^{O(c\log c)}$ vertices whose 
graph language is $\lang_G^c$, i.e., the set  of DAGs with global slice width at most $c$. 
\end{lemma}
\begin{proof}
In order to construct $\slicegraph^c$ , we create one vertex $\mathfrak{v}_{\boldS,f}$ for each unit slice $\boldS$ of width 
at most $c$, and each $c$-flow coloring $f$ of $\boldS$. We label the vertex $\mathfrak{v}_{\boldS,f}$ with the slice $\boldS$.
A slice $\boldS$ of width $c$ has at most $2c$ edges. Since in a $c$-flow coloring of $\boldS$, each edge receives a value 
between $1$ and $c$, there exist at most $c^{2c}=O(2^{O(c\log c)})$ ways of coloring $\boldS$. Since there are at most $2^{O(c\log c)}$ 
unit slices of width at most $c$, then $\slicegraph^c$ will still have at most $O(2^{O(c\log c)})$ vertices. Now we connect a 
vertex $\mathfrak{v}_{\boldS,f}$ to the vertex $\mathfrak{v}_{\boldS',f'}$ if and only if $\boldS$ can be glued to $\boldS'$
and if the values associated by $f$ to the out-frontier edges of $\boldS$ agree with the values associated by $f'$ to their respective edges
touching the in-frontier of $\boldS'$. By this construction a unit decomposition $\boldS_1\boldS_2...\boldS_n$ of a graph $G$ has a $c$-flow coloring
$f_1f_2...f_n$ if and only if there is an accepting walk $\mathfrak{v}_{\boldS_1,f_1}\mathfrak{v}_{\boldS_2,f_2}...\mathfrak{v}_{\boldS_n,f_n}$ in $\slicegraph^c$
such that $\boldS_1\circ\boldS_2\circ ... \circ \boldS_n = G$. By Corollary \ref{corollary:SlicedCharacterizationGlobalSliceWidth}, $G$ has global slice 
width at most $c$. 
\end{proof}

\begin{lemma}[Globally Bounded Hasse Diagram Generator]
\label{lemma:UniversalHasseDiagramGenerator}
For each $c\in \N$ with $c\geq 1$ there is a saturated Hasse diagram generator $\hasseGenerator^c$ on $2^{O(c^2)}$ 
vertices whose partial order language is $\lang^c_{PO}$, i.e., the set of partial 
orders whose Hasse diagrams have global slice width at most $c$.
\end{lemma}
\begin{proof}
As a first step, we construct slice graph $\slicegraph^c$ of Lemma \ref{lemma:ConstructionGlobalSliceWidth} which generates 
precisely the set of DAGs of global slice width at most $c$, and has $2^{O(c\log c)}$ vertices. Subsequently, we apply our transitive reduction algorithm 
(Theorem \ref{theorem:TransitiveReductionSliceGraphsB}) to obtain a Hasse diagram generator $\hasseGenerator^c$ on at most 
$2^{O(c\log c)}\cdot 2^{O(c^2)} = 2^{O(c^2)}$ vertices representing 
the same partial order language. By Proposition \ref{proposition:PreservesSaturation}, $\hasseGenerator^c$ is weakly saturated, and 
thus by Theorem \ref{theorem:WeakSaturation} it can be transformed into a fully saturated HDG. 
$\square$
\end{proof}

\subsection{Decidability, Closures and Canonization}
\label{subsection:DecidabilityClosuresCanonization}
\label{subsection:CanonizationSliceGraphs}

In this subsection we state closure, decidability and canonizability properties for the class of globally 
bounded DAG languages (Lemma \ref{lemma:ClosuresGraphLanguage}) and for the class of globally 
bounded partial order languages that can be represented by saturated slice languages (Theorem \ref{theorem:ClosuresPartialOrderLanguage}).

A function $\canonical_G$ canonizes slice graphs w.r.t. the graph language they generate if 
$i$) $\lang_{G}(\canonical_G(\slicegraph))=\lang_{G}(\slicegraph)$ and 
$ii$) for every slice graphs $\slicegraph, \slicegraph'$, $\canonical_G(\slicegraph)$ is
isomorphic to $\canonical_G(\slicegraph')$ precisely when $\lang_{G}(\slicegraph)=\lang_{G}(\slicegraph')$.
Similarly, a function $\canonical_{PO}$ canonizes slice graphs w.r.t. the partial order
language they generate if $i$) $\lang_{PO}(\canonical_{PO}(\slicegraph))=\lang_{PO}(\slicegraph)$ 
and $ii)$ for every slice graphs $\slicegraph, \slicegraph'$, $\canonical_{PO}(\slicegraph)$ is
isomorphic to $\canonical_{PO}(\slicegraph')$ precisely when $\lang_{PO}(\slicegraph)=\lang_{PO}(\slicegraph')$. 
We notice that it is hopeless to try to devise a canonization algorithm that works for every slice graph both with respect to 
their graph languages and with respect to their partial order languages. For instance, if we were able to 
compute canonical forms for graph languages $\lang_G, \lang_G'$ represented by general slice graphs, we would be able 
to decide $\lang_{G}\subseteq \lang'_{G}$ by testing whether $\lang_{G} \cup \lang_{G}' = \lang_{G}'$ \footnote{Clearly $\lang''=\lang
\cup \lang'$ if and only if $\lang''_{G}=\lang_{G} \cup \lang_{G}' $.}. However, inclusion of the graph 
languages generated by slice graphs is known to be undecidable \cite{deOliveiraOliveira2010}. 
Fortunately, as stated in Lemma \ref{lemma:ClosuresGraphLanguage} and in Theorem \ref{theorem:ClosuresPartialOrderLanguage},
such canonizability results are accomplishable for the class of saturated slice graphs. 

\begin{lemma}[DAG languages: Computability, Decidability and Canonization] 
\label{lemma:ClosuresGraphLanguage}
Let $\slicegraph$ and $\slicegraph'$ be two slice graphs over the alphabet $\slicealphabet^c$ generating slice languages $\lang$ and $\lang'$ respectively, 
and suppose $\slicegraph$ is saturated.
Then  
\begin{enumerate} 
\item one may compute 
	\begin{itemize}
		\item a slice graph $\slicegraph^{\cup}_G$ whose graph language is $\lang_{G} \cup \lang'_{G}$, 
		\item a slice graph $\slicegraph^{\cap}_G$ whose graph language is  $\lang_{G} \cap\lang'_{G}$ and,
		\item a slice graph $\overline{\slicegraph}_G$ whose  graph language is $\overline{\lang_{G}}\cap \lang_G^c$. 
	\end{itemize}
	furthermore, if $\slicegraph'$ is also saturated then so are $\slicegraph^{\cup}$, $\slicegraph^{\cap}$.
\item one may decide
	\begin{itemize}
		\item whether $\lang'_G \subseteq \lang_{G}$ and,
		\item whether $\lang_G \cap \lang'_{G} = \emptyset$. 
	\end{itemize}
\item one may compute a canonical saturated slice graph $\canonical_G(\slicegraph)$ generating $\lang_G$. 
\end{enumerate}
\end{lemma}
\begin{proof}
Since $\lang$ is saturated, equation \ref{equation:SaturatedUnion} implies that $\lang_{G} \cup \lang'_{G}$ 
iff $\lang \cup \lang'$, $\lang_{G} \cap \lang'_{G}$ iff $\lang \cap \lang'$, $\lang_G' \subseteq \lang_G$ iff
$\lang' \subseteq \lang$ and $\lang_G\cap \lang_G' = \emptyset$ iff $\lang\cap \lang'=\emptyset$, while $\overline{\slicegraph}$
is the slice graph whose slice language is $\lang(\slicegraph^c)\backslash \lang$, where $\slicegraph^c$
is the slice graph constructed in Lemma \ref{lemma:ConstructionGlobalSliceWidth}. Since it 
is well known that regular languages are closed under union, intersection and complementation, and since 
$\lang$, $\lang'$ and $\lang(\slicegraph^c)$ are regular subsets of $(\slicealphabet^c)^*$, items $1$ and $2$ follow.
Also, regular language theory says that there is a minimal canonical deterministic finite automaton $\automaton$ 
over $\slicealphabet^c$ generating $\lang$. By fixing a function $h$ that maps automata to 
labeled graphs representing the same regular language (e.g. see Appendix of \cite{deOliveiraOliveira2010}),
we may set the canonical form $\canonical(\slicegraph)$ to be the slice graph $h(\automaton)$. Since $\lang$ is 
saturated, $h(\automaton)$ will also be a canonical representative for the graph language $\lang_G$. 
$\square$.  
\end{proof}

As noted in Section \ref{section:Consequences}, there exist ({\bf even saturated}) regular slice languages $\lang$ and $\lang'$ for which 
$\lang_{G} \cap \lang'_{G} = \emptyset$ but $\lang_{PO} \cap \lang'_{PO} \neq \emptyset$, 
or for which $\lang_{G} \nsubseteq \lang'_{G}$ but $\lang_{PO} \subseteq \lang'_{PO}$. 
For instance, consider the "diamond" graph $D$ of Figure \ref{figure:langNotSaturated}.$iii$, and a graph $D'$ obtained from $D$ by adding 
an edge from its minimal to its maximal vertex. Then the languages $\lang=ud(D)$ and $\lang'=ud(D')$ consisting of 
all unit decompositions of $D$ and $D'$ respectively, are saturated slice languages. However both $\lang \cap \lang'=\emptyset$ and $\lang_G\cap\lang_G'=\emptyset$, 
while $\lang_{PO}\cap\lang'_{PO}$ is not empty, since $D$ and $D'$ induce the same partial order. 
Thus, as it will be clear in the proof of the next theorem (Theorem \ref{theorem:ClosuresPartialOrderLanguage}), our transitive reduction algorithm is 
essential for the statement of decidability and computability results concerning the partial order languages represented by slice graphs.

\begin{theorem}[Partial Order Languages: Computability, Decidability and Canonization]
\label{theorem:ClosuresPartialOrderLanguage}
Let $\slicegraph$ and $\slicegraph'$ be two slice graphs over the alphabet $\slicealphabet^c$ generating slice languages
$\lang$ and $\lang'$ respectively, and suppose $\slicegraph$ is saturated. Then  
\begin{enumerate} 
\item one may compute 
	\begin{itemize}
		\item a slice graph $\slicegraph^{\cup}_{PO}$ whose partial order language is $\lang_{PO} \cup \lang'_{PO}$, 
		\item a slice graph $\slicegraph^{\cap}_{PO}$ whose partial order language is  $\lang_{PO} \cap\lang'_{PO}$ and,
		\item a saturated slice graph $\overline{\slicegraph}_{PO}$ whose partial order language is $\overline{\lang}_{PO} \cap \lang_{PO}^c$. 
	\end{itemize}
	furthermore, if $\slicegraph'$ is also saturated then so are $\slicegraph^{\cup}_{PO}$ and $\slicegraph^{\cap}_{PO}$.
\item one may decide
	\begin{itemize}
		\item whether $\lang_{PO}(\slicegraph') \subseteq \lang_{PO}(\slicegraph)$ and,
		\item whether $\lang_{PO}(\slicegraph) \cap \lang_{PO}(\slicegraph') = \emptyset$. 
	\end{itemize}
\item one may compute a canonical saturated Hasse diagram generator $\canonical_{PO}(\slicegraph)$ generating $\lang_{PO}$. 
\end{enumerate}
\end{theorem}
\begin{proof}
As a crucial step towards all the results stated in the present theorem, we apply our transitive reduction algorithm to 
both $\slicegraph$ and $\slicegraph'$ (Theorem \ref{theorem:TransitiveReductionSliceGraphsB}), obtaining in this 
way a Hasse diagram generator $\hasseGenerator$ and $\hasseGenerator'$ representing the same partial order languages as 
$\slicegraph$ and $\slicegraph'$ respectively. The cruciality of this step stems from the fact that 
several ({\bf even saturated}) slice graphs may represent the same partial order language. 
Proposition \ref{proposition:PreservesSaturation} guarantees that $\hasseGenerator$ is weak 
saturated, and thus it can be transformed into a fully saturated HDG by Theorem \ref{theorem:WeakSaturation}.
Since the graph languages $\lang_G$ and $\lang'_G$ generated by $\hasseGenerator$ and $\hasseGenerator'$ respectively 
are transitive reduced, their partial order languages are in a bijective correspondence with their respective graph 
languages, and thus $\lang_{PO}\cup \lang'_{PO}$ iff $\lang_{G} \cup \lang'_{G}$, 
$\lang_{PO}\cap\lang'_{PO}$ iff $\lang_{G} \cap\lang'_{G}$, 
$\lang'_{PO} \subseteq \lang_{PO}$ iff $\lang'_{G} \subseteq \lang_{G}$, 
and $\lang_{PO} \cap \lang'_{PO} = \emptyset$ iff $\lang_{G} \cap \lang'_{G} = \emptyset$. Thus $\slicegraph^{\cup}_{PO}=\hasseGenerator^{\cup}_{G}$, 
$\slicegraph^{\cap}_{PO}=\hasseGenerator^{\cap}_{G}$ and the canonical form $\canonical_{PO}(\slicegraph) = \canonical_G(\hasseGenerator)$ 
can be computed by using Lemma \ref{lemma:ClosuresGraphLanguage}. 
Similarly inclusion and emptiness of intersection can be decided by applying Lemma \ref{lemma:ClosuresGraphLanguage}. In order to compute $\overline{\slicegraph}_{PO}$, 
instead of applying Lemma \ref{lemma:ClosuresGraphLanguage} we set $\overline{\slicegraph}_{PO}$ to be the Hasse diagram generator whose slice language is 
$\lang(\hasseGenerator^c)\backslash \lang$ where $\hasseGenerator^c$ is the Hasse diagram generator 
constructed in Lemma \ref{lemma:UniversalHasseDiagramGenerator}.
$\square$
\end{proof}

\subsection{Weak Saturation, Saturation and Loop Connectivity}
\label{subsection:WeakSaturationLoopConnectivity}

We observe that in general it is not possible to effectively transform a non-saturated slice
graph $\slicegraph$ into a saturated slice graph $\slicegraph'$ representing
the same partial order language, since this would imply canonization of
arbitrary slice graphs (see Section \ref{subsection:CanonizationSliceGraphs}). 
Indeed from results of \cite{Sakarovitch1992} and from our view of saturated slice languages over 
an alphabet $\slicealphabet^c$ as being closed under a commutation operation on $\slicealphabet^c$, 
we can conclude that even determining whether there exists a saturated slice graph $\slicegraph'$ representing 
the same partial order language as $\slicegraph$ is undecidable. 

In this subsection we prove that weakly saturated slice graphs can be transformed
into saturated slice graphs representing the same partial order language. From this 
result we conclude that all decidability results that are valid for regular saturated 
slice languages with regard to the partial order language they generate are equally 
valid for weakly saturated slice languages. We also 
introduce the concept of {\em loop-connectivity}, which is a topological property of 
slice graphs. Slice graphs satisfying this property can also be saturated. Both weak
saturation and loop-connectivity will have applications to concurrency theory, in the 
sense that recognizable trace languages \cite{HussonMorin2000} can be mapped to weakly saturated slice languages, 
while linearization regular MSC languages generated by message sequence chart graphs \cite{HenriksenMukundKumarSohoniThiagarajan2005}
can be mapped to loop-connected slice graphs.

\begin{theorem} \label{theorem:WeakSaturation} Let $\slicegraph$ be a 
weakly saturated slicegraph over $\slicealphabet^c$, and suppose that $\slicegraph$ 
has $n$ vertices. Then there exists a saturated slicegraph $\slicegraph'$ on 
$n\cdot O(2^c)$ vertices generating the same partial order language, i.e., such that
$\lang_{PO}(\slicegraph)=\lang_{PO}(\slicegraph')$.  \end{theorem}
\begin{proof}
We write $\mathfrak{S}_n$ for the symmetric group on $n$ elements. Let $\boldS$
be a unit slice with in-frontier $I$ and out-frontier $O$, $\pi$ be a
permutation in $\mathfrak{S}_{|I|}$ and $\sigma$ be a permutation in
$\mathfrak{S}_{|O|}$. We write $(\boldS,\pi,\sigma)$ for the unit slice
obtained from $\boldS$ by permuting the labels of the in-frontier vertices
according to $\pi$ and the labels of the out-frontier vertices according to
$\sigma$. The saturated version of $\slicegraph$ is obtained by replacing each
vertex $v$ in $\mathcal{V}$ by a set of vertices $\{v_{\pi\sigma}\}$ and
labeling each $v_{\pi,\sigma}$ with the slice $(\mathcal{S}(v),\pi,\sigma)$.
We add an edge $(v_{\pi\sigma}, v_{\pi'\sigma'})$ to $\mathcal{E}'$ if and
only if there is an edge from $v$ to $v'$ in $\slicegraph$ and if
$\sigma=\pi'$.  We note that $\lang_G(\slicegraph)=\lang_G(\slicegraph')$.
Also, since $(\boldS,\pi,\sigma)\circ (\boldS',\pi',\sigma') =
(\boldS\circ\boldS', \pi, \sigma')$ whenever $\sigma=\pi'$, we can see that
each two unit decompositions $\boldS_1\boldS_2\cdots\boldS_n$ and
$\boldS'_1\boldS'_2\cdots\boldS'_n$ of a DAG $H$ corresponding to the same
topological ordering $u_1u_2\cdots u_n$ of its vertices, are related by
permutations of the labels of the frontier vertices of $\boldS_i$.  Since
$\slicegraph$ is weakly saturated the slice language of the new slice graph
$\slicegraph'$ contains the whole set of unit decompositions $ud(H)$ of each
$H$ in $\lang_{G}(\slicegraph)$. $\square$ 
\end{proof}

We recall that a directed graph is strongly connected if for any two vertices $v$ and 
$v'$ there is a path going from $v$ to $v'$ and a path from $v'$ to $v$. Below we will 
define the notion of loop-connected slice graph. In Theorem \ref{theorem:LoopConnected}
we will prove that every loop-connected slice graph can be transformed into a saturated 
slice graph representing the same set of partial orders.

\begin{definition}[Loop-Connected Slice Graph]
\label{definition:LoopConnected}
A slice graph $\slicegraph=(\mathcal{V},\mathcal{E},\slice)$ is loop connected if for every loop 
$\mathfrak{v}_1\mathfrak{v}_2...\mathfrak{v}_n\mathfrak{v_1}$ in $\slicegraph$ the graph obtained by gluing
the out-frontier of the slice $\slice(\mathfrak{v}_1)\circ \slice(\mathfrak{v}_2)\circ...\circ\slice(\mathfrak{v}_n)$
with its own in-frontier has a unique strongly connected component (Fig. \ref{figure:notMazurkiewicz}.$ii$).
\end{definition}

\begin{theorem}
\label{theorem:LoopConnected}
For every loop-connected slice graph $\slicegraph$ there is a saturated slice graph $\slicegraph'$
representing the same partial order language. 
\end{theorem}
\begin{proof}
Let $\sliceautomaton$ be the minimal deterministic automaton over the slice alphabet $\Sigma_{\slice}$ which generates the same 
slice language as $\slicegraph$, and let $n$ be the number of states in $\sliceautomaton$. We repeat the following procedure
$n$ times: For each path $q_1\stackrel{\boldS_1}{\longrightarrow} q_2 \stackrel{\boldS_2}{\longrightarrow}q_3$ and 
each two slices $\boldS_1'$ and $\boldS_2'$ such that $\boldS_1\circ\boldS_2 = \boldS_1'\circ\boldS_2'$, add a state 
$q_2'$  to the automaton and the transitions $q_1\stackrel{\boldS_1'}{\longrightarrow}q_2'$ and $q_2'\stackrel{\boldS_2'}{\longrightarrow}q_3$
if such a state is not already present in the automaton. 
We claim that if $\slicegraph$ is loop-connected then after iterating this step $n$ times, $\sliceautomaton$ 
will generate a saturated slice language representing the same set of partial orders. To see this, let $\sliceautomaton^n$ be the automaton 
after the $n$-th iteration and suppose it is not saturated. Then for some slice string $\boldS_1\boldS_2...\boldS_m$ in $\lang(\sliceautomaton^n)$ with
$m>n$, there exists $i$ such that $\boldS_i$ is independent of $\boldS_{i+1}$ and there are $\boldS_i'$ and $\boldS_{i+1}'$ such that 
$\boldS_i\circ\boldS_{i+1}= \boldS_i'\circ\boldS_{i+1}'$ but $\boldS_1\boldS_2...\boldS_i'\boldS_{i+1}'...\boldS_m$ is not 
in $\lang(\sliceautomaton^n)$. This means that for some slice string $\boldS_1''\boldS_2''...\boldS_m''$ in the slice language of 
the original slice automaton $\sliceautomaton$, and for some $i,j$ with $j-i> n$ there is no path from the center vertex of 
$\boldS_i''$ to the center vertex of $\boldS_j''$ in the composed slice $\boldS_1''\circ \boldS_2'' \circ ...\circ \boldS_m''$. 
From the pumping lemma for regular languages we know that there exist slice strings $x,y,z \in \slicealphabet^*$ such that 
$\boldS_i''\boldS_{i+1}''...\boldS_j'' = xyz$ and such that $xy^rz\in \lang(\sliceautomaton)$ for every $r\geq 0$ and thus the slice string 
$y$ labels a cycle in $\sliceautomaton$. Since there is no path from the center vertex of $\boldS_i$ to the center vertex of 
$\boldS_j$ then gluing the in-frontier of the slice $\slice(y)$ with its own out-frontier, we have a graph which is not 
strongly connected.$\square$ 
\end{proof}

\section{Mazurkiewicz Traces, Message Sequence Chart Languages, and Slice Graphs} 
\label{section:Reductions}

In this section we show how to describe two well known formalisms 
used in concurrency theory in terms of slices. Namely, we 
will show that partial order languages represented 
through Mazurkiewicz traces or through message-sequence-chart languages 
can also be represented by slice graphs. We emphasize that slice graphs
that arise from natural reductions may fall short of being transitive reduced. This observation 
illustrates the fact that general slice graphs may be substantially easier
to reason about at a preliminary stage of specification when compared with Hasse diagram 
generators. It also illustrates one more application of our transitive reduction algorithm:
by transitive reducing these slice graphs, and applying the results of \cite{deOliveiraOliveira2010}
we may use Mazurkiewicz traces and MSC-languages as a point of departure for the verification and 
synthesis of $p/t$-nets. This will be the topic of next section (Section \ref{section:PetriNets}).

\subsection{Mazurkiewicz Traces}
\label{subsection:MazurkiewiczTraces}

In Mazurkiewicz trace theory, partial orders are represented as equivalence
classes of words over an alphabet of events \cite{Mazurkiewicz1986}. Given an
alphabet $\Sigma$ of events and a symmetric and anti-reflexive {\em
independence relation} $I\subseteq \Sigma\times \Sigma$, a string $\alpha
ab\beta$ is defined to be similar to the string $\alpha ba \beta$ ($\alpha ab
\beta \simeq \alpha ba \beta$) if $aIb$. A trace is then an equivalence class
of the transitive and reflexive closure of $\simeq^*$ of the relation $\simeq$.
We denote by $[\alpha]_{I}$ the trace corresponding to a string $\alpha\in
\Sigma^*$. 
A partial order $po_I(\alpha)$ is associated with a string $\alpha\in \Sigma^*$
of events in the following way: First we consider a dependence DAG
$dep_I(\alpha)=(V,E,l)$ that has one vertex $v_i\in V$ labeled by the event
$\alpha_i$ for each $i\in\{1, ... ,|\alpha|\}$.  An edge connects $v_i$ to
$v_j$ in $E$ if and only if $i<j$ and $(\alpha_i,\alpha_j)\notin I$. Then $po_I(\alpha)$
is the transitive closure of $dep_I(\alpha)$.  One may verify that two strings
induce the same partial order if and only if they belong to the same trace. The
trace language induced by a string language $\lang\subseteq \Sigma^*$ with
respect to an independence relation $I$ is the set $[\lang]_I = \{[\alpha]_{I}|
\alpha \in \lang\}$ and the trace closure of $\lang$ is the language $\lang^I =
\cup_{\alpha \in \lang} [\alpha]$. Given a finite automaton $\automaton$ over
an alphabet $\Sigma$ and an independence relation $I\subset \Sigma\times
\Sigma$, we denote by $\lang(\automaton)$ the regular language defined by
$\automaton $ and by $\lang_{PO}(\automaton,I) = \{po_I(\alpha)| \alpha \in
\lang(\automaton)\}$ the partial order language induced by $(\automaton,I)$.
The next lemma (Lemma \ref{lemma:TracesToSlices}) says that for any finite
automaton $\automaton$ and independence relation $I$, there is a slice graph
$\slicegraph(\automaton,I)$ inducing the same partial order language as
$(\automaton,I)$. We notice again that $\slicegraph(\automaton,I)$ is not at all 
guaranteed to be a Hasse diagram generator. 

\begin{lemma}[From Traces to Slices] \label{lemma:TracesToSlices} Let
$\automaton$ be a finite automaton over an alphabet $\Sigma$ and $I\subset
\Sigma\times\Sigma$ an independence relation. Then there exists an effectively
constructible slice graph $\slicegraph(\automaton,I)$ such that
$\lang_{PO}(\automaton,I) = \lang_{PO}(\slicegraph(\automaton,I))$.
Furthermore, if $\lang(\automaton)$ is trace closed, then $\slicegraph$ is
weakly saturated.  \end{lemma} \begin{proof} From an independence alphabet
$(\Sigma,I)$ we will derive a slice alphabet $\Sigma_{\slice}=\{\boldS_a| a\in
\Sigma\}$ (FIG. \ref{figure:TracesToSlices}.$vi$) in such a way that the
partial order $po_I(\alpha)$ induced by a string $\alpha=\alpha_1\alpha_2
\cdots \alpha_k \in \Sigma^*$ will be identical to the partial order induced by
the slice string $\boldS_{\alpha_1}\boldS_{\alpha_2}\cdots\boldS_{\alpha_k}$
over $\Sigma_{\slice}$.  In other words, $po_I(\alpha)$ will be equal to the
transitive closure of the DAG
$\boldS(\alpha)=\boldS_1\circ\boldS_2\circ...\circ\boldS_k$.  We assume without
loss of generality that $\Sigma$ has two special symbols $\iota$ and
$\varepsilon$ that are not independent from any other symbol in $\Sigma$. The
initial symbol $\iota$ appears a unique time in the beginning of each word
accepted by $\automaton$ while the final symbol $\varepsilon$ appears a unique
time at the end of each word. Let $\Sigma'=\Sigma\backslash
\{\iota,\varepsilon\}$, and $D= (\Sigma'\times \Sigma')\backslash I$ be a
dependence relation. For each symbol $a\in \Sigma$ we define the slice
$\boldS_a$ as follows: Both the in-fronter $I$ and the out-frontier $O$ of
$\boldS_a$ have $|D|$ vertices indexed by $D$, and the center of $\boldS_a$ has
a unique vertex  $v_a$ which is labeled by $a$. In symbols $I=\{ I_{ab} |
\{a,b\} \in D\}$ and $O=\{ O_{ab} | \{a,b\} \in D\}$.  For each pair $\{b,c\}
\in D$ with $a\neq b$ and $a\neq c$ we add an edge $(I_{bc}, O_{bc})$ in
$\boldS_a$, and for each pair $\{a,x\} \in D$ we add edges $(I_{ax}, v_a)$ and
$(v_a, O_{ax})$ into $\boldS_a$ (FIG. \ref{figure:TracesToSlices}.$vi$). We
associate with $\iota$ an initial slice $\boldS_{\iota}$, with center vertex
$v_{\iota}$ and out-frontier $O$, and to $\varepsilon$, a final slice
$\boldS_{\varepsilon}$ with center vertex $v_{\varepsilon}$ and in-frontier
$I$.  We may assume that the DAGs $\boldS(\alpha)= \boldS_{\alpha_1}\circ
\boldS_{\alpha_2} \circ ...\circ \boldS_{\alpha_k}$ and dependence DAG
$dep_I(\alpha)$ associated with a string $\alpha\in \Sigma^*$ have identical
sets of vertices, with vertex $v_i$ corresponding to the $i$-th symbol of
$\alpha$. Nevertheless these DAGs are not isomorphic. Neither one is
necessarily a subgraph of the other. However one can verify the following fact:
for each edge $(v_i,v_j) \in dep_I(\alpha)$ there is a path in $\boldS(\alpha)$
joining vertices $v_i$ to $v_j$. Conversely, for each edge $(v_i,v_j)$ in
$\boldS(\alpha)$ there is a path joining $v_i$ to $v_j$ in $dep_I(\alpha)$.
Hence, both $\boldS(\alpha)$ and $dep_{I}(\alpha)$ induce the same partial
order.  Let $f:\Sigma\rightarrow \Sigma_{\slice}$ be the isomorphism that maps
each symbol $a\in \Sigma$ to its slice $\boldS_a$. Then the isomorphic image of
$\lang(\automaton)$ under $f$ is a regular\footnote{The term {\em regular} here is
used in a fair sense, since $f$ maps isomorphically the free monoid generated
by $\Sigma$ to the free monoid generated by $\Sigma_{\slice}$.} slice language
inducing $\lang_{PO}(\automaton,I)$, and thus can be represented by a slice
graph $\slicegraph(\automaton,I)$.  
$\square$ 
\end{proof}

There is a substantial difference between our notion of independence, defined
on slice alphabets and the notion of independence in Mazurkiewicz trace theory.
While the independence relation on slices is determined solely with basis on
the structure of the slices (Fig. \ref{figure:mazurkiewicz}.$iii$), without
taking into consideration the events that label their center vertices, the
Mazurkiewicz independence relation is defined directly on events.  As a
consequence, once an independence relation $I$ is fixed, the nature of the
partial orders that can be represented as traces with respect to $I$ is
restricted. This is valid even for more general notions of traces, such as
Diekert's semi-traces \cite{Diekert1994} and the context dependent traces of
\cite{HoogersKleijnThiagarajan1995}, in which for instance, partial orders
containing auto-concurrency cannot be represented. In our setting any partial
order $po$ labeled over a set of events $T$ may be represented by a slice
trace: namely the set of unit decompositions of its the Hasse diagram.

\subsection{From MSC-Languages to Slice Graphs}
\label{subsection:MSCLanguages}

\begin{figure}[h] \centering \includegraphics[scale=0.35]{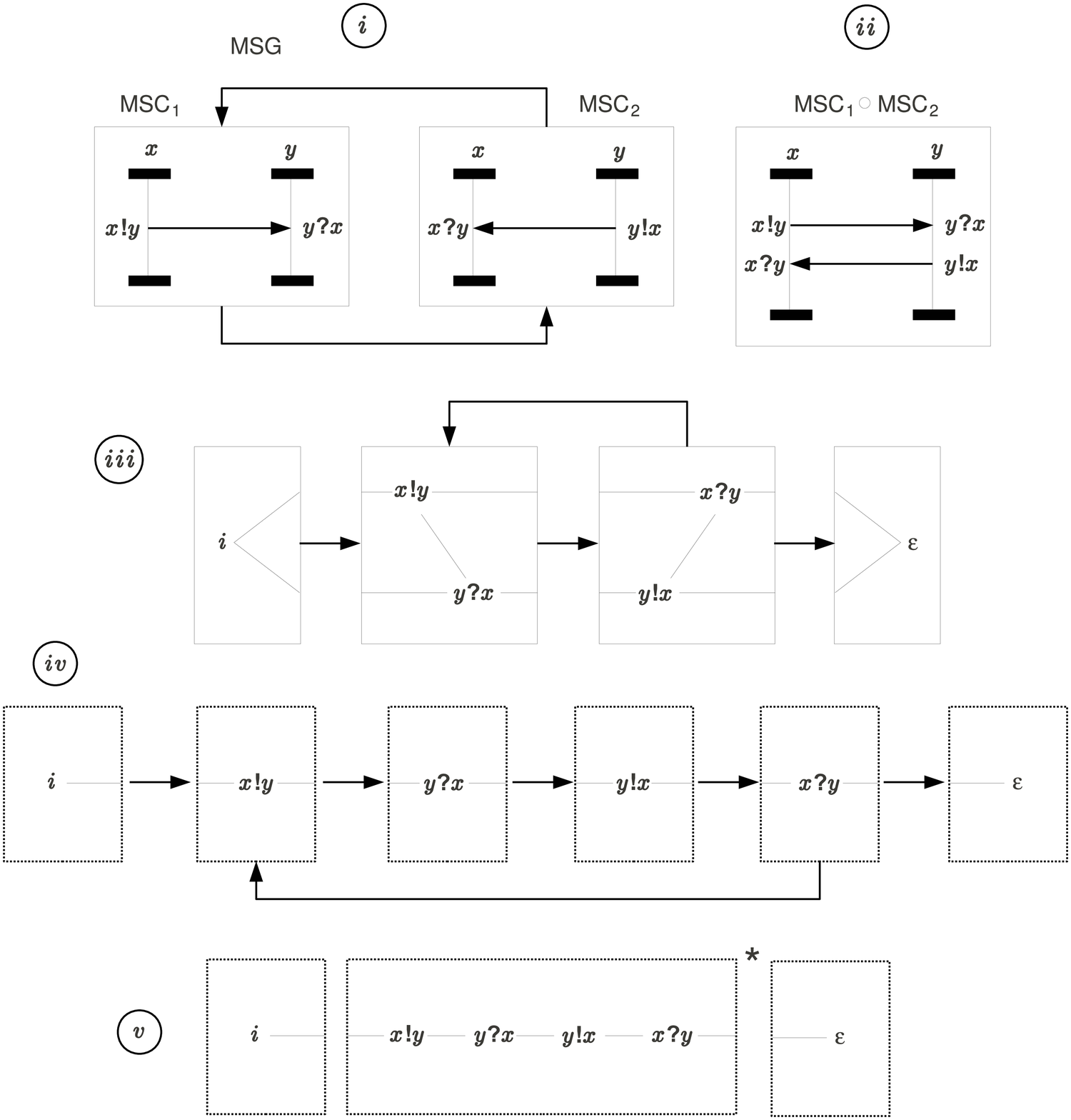}
\caption{$i$) MSG whose vertices are labeled with $MSC_1$ and $MSC_2$. $ii$) 
Composition of $\mbox{MSC}_1\circ \mbox{MSC}_2$. $III$) Direct translation from $MSG$ to
a slice graph which is not transitive reduced. $iv$) Canonical slice 
graph generating the same set of partial orders. $v$) An elegant way of 
specifying the same language using a slice expression. 
\label{figure:MSCToSlices}}\end{figure}

Message Sequence Charts (MSCs) are used to depict the exchange of messages between 
the processes of a distributed system along a single partially ordered execution. Although 
being only able to represent partial orders of a very special type, MSCs find several 
applications and are in special suitable to describe the behavior of telecommunication protocols. 
Infinite families of MSCs can be specified by hierarchical (or high-level) message sequence
charts (HMSCs) or equivalently, by message sequence graphs (MSGs) \cite{AlurYannakakis1999,MuschollPeledSu1998,Morin2001}. 
In this section we chose to work with message sequence graphs for 
they have a straightforward analogy with slice graphs. Namely, 
message sequence graphs are directed graphs without multiple edges, 
but possibly containing self loops, whose vertices are labeled with MSCs instead of with slices.
Thus our translation from MSGs to slice graphs amounts to translate MSCs to 
slices in such a way that the composition of the former yields the same partial 
orders as the composition of the latter. We notice that the resulting slice graphs 
are not guaranteed to be transitive reduced. However, by using our transitive reduction 
algorithm, they can be further reduced in into Hasse diagram generators (Fig. \ref{figure:MSCToSlices}).

We formalize MSCs according to the terminology in \cite{Morin2001}. 
Let $\processes$ be a finite set of processes, also called instances. For any instance 
$i\in \mathcal{I}$, the set $\Sigma^{int}_i$ denotes a finite set of 
{\em internal actions},  $\Sigma^{!}_{i}=\{i!j| j\in \processes\backslash\{i\}\}$ a 
set of {\em send actions} and $\Sigma^{?}_{i}=\{i?j|j\in \processes\backslash \{i\}\}$
a set of  {\em receive actions}. The alphabet of events associated with 
the instance $i$ is the disjoint union of these three sets: 
$\Sigma_i=\Sigma_i^{int} \cup \Sigma_i^{!} \cup \Sigma_i^{?}$. We shall assume
that the alphabets $\Sigma_i$ are disjoint and let 
$\Sigma_{\processes} = \bigcup_{i\in \processes} \Sigma_i$. 
Given an action $a\in \Sigma_{\processes}$, $Ins(a)$ denotes the unique instance 
$i$ such that $a\in \Sigma_i$. Finally, for any partial order $po=(V,E,l)$
whose vertices are labeled over $\Sigma_{\mathcal{I}}$, we denote by $Ins(v)$
the instance on which the event $v\in V$ occurs: $Ins(v)=Ins(l(v))$. 

\begin{definition}[Message Sequence Chart (MSC)]
A message sequence chart is a partial order $M=(V,\leq,l)$ 
over $\Sigma_{\processes}$ such that
\begin{itemize}
	\item Events occurring on the same process are linearly 
		ordered: For every pair of events $v,v'\in V$ if 
	$Ins(v)=Ins(v')$ then $v\leq v'$ or $v'\leq v$. 
	\item For any two distinct processes $i,j$, there are 
		as many send events from $i$ to $j$ as receive 
		events of $j$ from $i$: $\#^{i!j}(V)=\#^{j?i}(V)$. 
	\item The $n$-th message sent from $i$ to $j$ is received
		when the $n$-th event $j?i$ occurs, i.e., the channels
		are assumed to be FIFO. $l(v)=i!j$ and $l(v')=j?i$
		and $\#^{i!j}(\downarrow v) = \#^{j?i}(\downarrow v')$ then 
		$v\leq v'$. 
	\item If $v\prec v'$ and $Ins(v) \neq Ins(v')$ then $l(v)=i!j$,
		$l(v')=j?i$ and $\#^{i!j}(\downarrow v)$ is equal to 
		$\#^{j?i}(\downarrow v')$.
\end{itemize}
\end{definition}

The composition $M\circ M'$ of two MSCs $M=(V,\leq,l)$ and $M'=(V',\leq',l')$ can be defined directly into the partial 
order level as the transitive closure of the graph $$M\cup M' \cup \{(v,v')\in V\times V'| Ins(v)=Ins(v')\}.$$ The partial order language generated by a message sequence 
graph $\msg$ is the set $\lang_{PO}(\msg)$ of all partial orders obtained by the composition of sequences of 
MSCs which labels walks in $\msg$. A language $\lang$ of MSCs is linearization-regular \cite{HenriksenMukundKumarSohoniThiagarajan2005} if 
its set of linearizations $lin(\lang) = \cup_{M\in\lang} lin(M)$ is recognizable in the free 
monoid $\Sigma_{\mathcal{I}}^{*}$. The connectivity graph of a MSC $M$ is the graph $C(M)$ whose 
vertices are the instances of $M$ and there is an edge from instance $i$ to instance $j$ if 
$i$ sends some message to $j$. An MSG $\msg=(\mathcal{V},\mathcal{E},\mathfrak{l})$ is locally synchronized \cite{MuschollPeled1999} (called bounded in \cite{AlurYannakakis1999}) 
if for each loop $w=v_1v_2...v_nv_1$ in $\msg$ the connectivity graph of the MSC $\mathfrak{l}(v_1)\circ \mathfrak{l}(v_2)\circ...\circ \mathfrak{l}(v_n)\circ \mathfrak{l}(v_1)$ has a 
unique strongly connected component. It can be proved that an MSC language generated by a MSG is linearization-regular if and only 
if it is locally synchronized \cite{HenriksenMukundKumarSohoniThiagarajan2005}.
In the next lemma we prove that the partial order language of any MSG can be represented by a 
slice graph. Furthermore, locally synchronized MSGs correspond to loop-connected slice graphs, 
which can be saturated by Theorem \ref{theorem:LoopConnected}.

\begin{lemma}[From MSCs to Slices]
Let $\msg$ be a message sequence graph. Then there exists a slice graph $\slicegraph_{\mathcal{M}}$ satisfying 
$\lang_{PO}(\mathcal{M})=\lang_{PO}(\slicegraph_{\mathcal{M}})$. Furthermore if $\lang_{PO}(\mathcal{M})$ is linearization-regular 
then $\slicegraph_{\mathcal{M}}$ is loop-connected. 
\end{lemma}
\begin{proof}
We associate to each MSC $M$ a slice $\boldS_M$ in such a way that for each two MSCs $M_1$ and $M_2$, the partial order $M_1\cdot M_2$ 
is equal to the partial order induced by the transitive closure of $\boldS_{M_1}\circ\boldS_{M_2}$ (modulo the frontier vertices). 
Each frontier of $\boldS_M$ will have $|\mathcal{I}|$ nodes, one for each instance $i\in \mathcal{I}$.
If $M$ is a MSC then the slice $\boldS_M$ is the Hasse diagram of $M$ together with the new frontier vertices and some new edges which 
we describe as follows: If for some instance $i\in\mathcal{I}$, there is no vertex $v$ of $M$ such that $Ins(v)=i$, then we add an 
edge from the $i$-th in-frontier of $\boldS_M$ to the $i$-th out-frontier of $M$. For all the other instances in $\mathcal{I}$, add an 
edge from the $i$-th in-frontier vertex of $\boldS_M$ to the unique minimal vertex $v$ of $M$ satisfying $Ins(v)=i$, and an edge from the 
unique maximal vertex $v'$ of $M$ satisfying $Ins(v')=i$ to the $i$-th out-frontier vertex of $\boldS_M$. We observe that although 
each slice $\boldS_M$ is transitive reduced, the composition $\boldS_{M_1}\circ\boldS_{M_2}$ is not necessarily transitive reduced 
(Fig. \ref{figure:MSCToSlices}.$iii$). Now if the partial order language represented by $\msg$ is linearization-regular, then $\msg$ is locally 
synchronized. Furthermore each sequence $M_1M_2...M_nM_1$ of MSCs labeling a loop 
in $\msg$ corresponds to a sequence $\boldS_{M_1}\boldS_{M_2}...\boldS_{M_n}\boldS_{M_1}$ labeling a loop in $\slicegraph$. One can 
then verify that if the connectivity graph of $M_1\circ M_2\circ...\circ M_n$ has a unique strongly connected component, then 
gluing the out frontier of the slice $\boldS_{M_1}\circ\boldS_{M_2}\circ ...\circ \boldS_{M_n}$ with its own in frontier we also have 
a unique strongly connected component, and thus $\slicegraph$ is loop-connected. 
$\square$ 
\end{proof}

\subsubsection{Comparison between MSC languages and Slice languages}
\label{subsection:Comparison}

\begin{figure}[hf] \centering \includegraphics[scale=0.35]{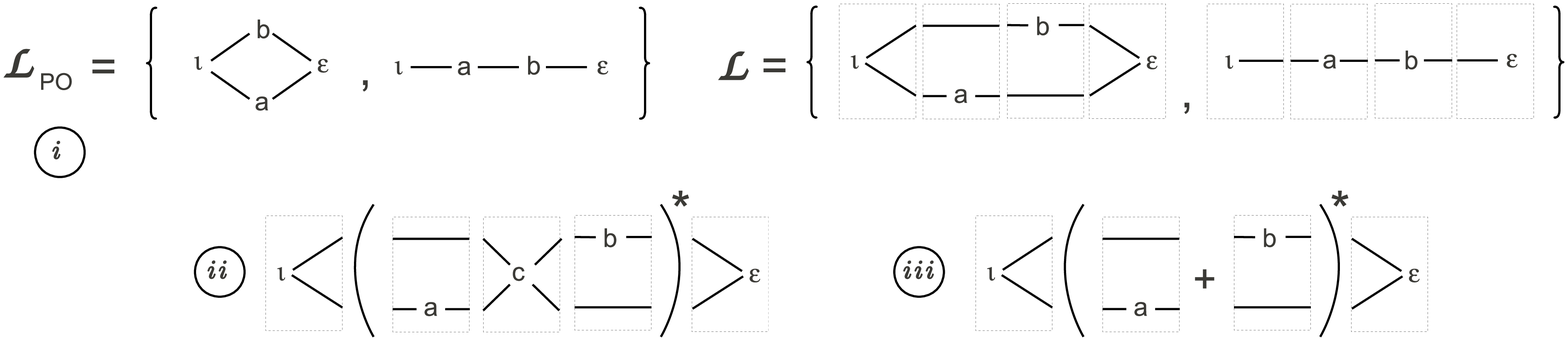}
\caption{  $i$) A simple partial order language $\lang_{PO}$ that cannot be represented through 
Mazurkiewicz traces, but which can be represented by a slice language $\lang$.
$ii)$ A slice expression whose corresponding slice graph is loop connected. 
$iii$) A slice expression whose corresponding slice graph is weakly saturated but not loop-connected.} 
\label{figure:notMazurkiewicz} 
\end{figure}

A special property which is satisfied by MSC's, and which is also 
observed in partial orders represented by Mazurkiewicz trances is the following: 
If $M,M'$ are two partial orders represented through MSC's (or through Mazurkiewicz traces) then $M$ is isomorphic to $M'$ if and only if 
$lin(M)\cap lin(M')=\emptyset$ \cite{Morin2001,HenriksenMukundKumarSohoniThiagarajan2005} where 
$lin(M)$ denotes the set of linearizations of $M$. 
This property which is fundamental for the development of several aspects of MSC-language theory and Mazurkiewicz trace theory, 
turns also to be a bottleneck for their expressiveness. For instance, some very simple partial order
languages, such as the one depicted in Figure \ref{figure:notMazurkiewicz}.$i$ cannot be represented by any formalisms satisfying this property. 
This bottleneck is not a issue when dealing with slices languages because the role of linearization of a partial order is 
completely replaced by the notion of unit decomposition of their Hasse diagrams.

A notion of atomic MSC has also been defined: An MSC $M$ is a component of a MSC 
$M'$ if there exist MSCs $M_1$ and $M_2$ such that $M'=M_1\circ M \circ M_2$.
$M$ is an atomic MSC if the only component of $M$ is $M$ itself.
Two atomic MSCs $M_1$ and $M_2$ if their vertices are labeled with actions from 
disjoint sets of processes (instances). We notice however that not every MSC can 
be decomposed into atomic MSC's consisting on a unique event, or more 
appropriately, 
consisting on a unique message being sent and received. Contrast this with the fact 
that any DAG can be written as a composition of unit slices.

 Finally, the notion of local synchronizability of message sequence graphs share some 
similarity with the notion of loop-connectivity of slice graphs. However, as mentioned above, 
an MSG $\msg$ generates a regular partial order language if and only if $\msg$ is locally synchronized,
while an analogous characterization of saturated slice graphs is not valid, i.e., while 
loop-connected slice graphs can be saturated, there are saturated slice graphs 
which are not loop-connected, as shown in Figure \ref{figure:notMazurkiewicz}.$iii$. 

\section{Applications to Petri Nets} 
\label{section:PetriNets}

We will start this section by providing a formal definition of $p/t$-nets and 
their partial order semantics. Subsequently we will give a simplified overview
of the main results in \cite{deOliveiraOliveira2010}, establishing whenever 
possible connections with the results proved in the previous sections. 
The main theorem of this section (Theorem \ref{theorem:RefinedExpressibility})
states that the causal behavior of any bounded petri net can be represented
by a canonical saturated Hasse diagram generator. While in our previous 
work we proved that the partial order behavior of $p/t$-nets may be represented
via Hasse diagram generators, we did not prove that this could be done in 
a canonical way. Herein, building on our development of saturated slice languages
we will show that this is indeed possible. We will end this section by 
stating a corollary (Corollary \ref{corollary:MSCAndPetriNets}) connecting 
Mazurkiewicz trace languages and MSC-languages to $p/t$-nets. 

\subsection{Partial order Semantics of $p/t$-nets}
\label{subsection:PetriNets}
\vspace{10pt}

Let $T$ be a finite set of transitions. Then a place over $T$ is a triple
$p=(p_0,\putsp,\takesp)$ where $p_0$ denotes the initial number of tokens in
$p$ and $\putsp,\takesp:T\rightarrow \N$ are functions which denote the number
of tokens that a transition $t$ respectively puts in and takes from $p$. A
$p/t$-net over $T$ is a pair $N=(P,T)$ where $T$ is a set of transitions and
$P$ a finite multi-set of places over $T$. We assume through this paper that
for each transition $t\in T$, there exist places $p_1,p_2\in P$ for which
$\putsp_1(t) >  0$ and $\takesp_2(t) > 0$. A marking of $N$ is a function
$m:P\rightarrow \N$. A transition $t$ is enabled at marking $m$ if $m(p)\geq
\takesp(t)$ for each $p\in P$. The occurrence of an enabled transition at
marking $m$ gives rise to a new marking $m'$ defined as $m'(p)=m(p) -\takesp(t)
+ \putsp(t)$. The initial marking $m_0$ of $N$ is given by $m_0(p)=p_0$ for
each $p\in P$. A sequence of transitions $t_0t_1...t_{n-1}$ is an occurrence
sequence of $N$ if there exists a sequence of markings $m_0m_1...m_n$ such that
$t_i$ is enabled at $m_i$ and if $m_{i+1}$ is obtained by the firing of $t_i$
at marking $m_i$. A marking $m$ is legal if it is the result of the firing of
an occurrence sequence of $N$. A place $p$ of $N$ is $k$-safe if $m(p)\leq k$
for each legal marking $m$ of $N$. A net $N$ is $k$-safe if each of its places
is $k$-safe. $N$ is bounded if it is $k$-safe for some $k$. The union of two
$p/t$-nets $N_1=(P_1,T)$ and $N_2=(P_2,T)$ having a common set of transitions
$T$  is the $p/t$-net $N_1\cup N_2=(P_1 \dot{\cup} P_2, T)$.  We consider that
the multiplicity of a place $p$ in $P_1\dot{\cup} P_2$ is the sum of its
multiplicities in $P_1$ and in $P_2$. 

\begin{definition}[Process] \label{definition:Process} A process of a $p/t$-net
$N=(P,T)$ is a $DAG$ $\pi=(B\dot{\cup}V,F,\rho)$ where the vertex set
$B\dot{\cup} V$  is partitioned into a set of {\em conditions} $B$ and a set of
{\em events} $V$.  $F \subseteq (B\times V)\cup (V\times B)$ and $\rho:(B\cup
V)\rightarrow (P\cup T)\cup \{\iota,\epsilon\}$ are required to satisfy the
following conditions:  \begin{enumerate} \item \label{process:itemOne} $\pi$
has a unique minimal vertex $v_{\iota}\in V$ and a unique maximal vertex
$v_{\varepsilon}\in V$.  \item \label{process:itemTwo} Conditions are
unbranched: $\forall b\in B, |\{(b,v)\in F\}| = 1 = |\{(v,b)\in F\}|$.  \item
\label{process:itemThree} Places label conditions and transitions label events.
Minimal and maximal vertices have special labels.  $$\rho(B)\subseteq P
\hspace{1cm} \rho(V\backslash\{v_{\iota},v_{\varepsilon}\})\subseteq T
\hspace{1cm} \rho(v_{\iota})=\iota
\hspace{1cm}\rho(v_{\varepsilon})=\varepsilon$$ \item \label{process:itemFour}
If $\rho$ labels an event $v\in V\backslash\{v_{\iota}, v_{\varepsilon}\}$ with
a transition $t\in T$ then 
for each $p\in P$, $v$ has $\takesp(t)$ preconditions and $\putsp(t)$ postconditions labeled by $p$:  
$$|\{(b,v)\in F :\rho(b)\!=\!p\}|\!=\!\takesp(t) \hspace{1cm} |\{(v,b)\in F :\rho(b)\!=\!p\}|\!=\!\putsp(t)$$
\item \label{process:itemSix}
For each $p\in P$, $v_{\iota}$ has $p_0$ post-conditions labeled by $p$\;:
$|\{(v_{\iota},b): \rho(b)\!=\!p\}|\!=\!p_0$.  
\end{enumerate} 
\end{definition}

The only point our definition of process differs from the usual definition of
$p/t$-net process \cite{GoltzReisig1983} is the addition of a minimal event
$v_{\iota}$ which is labeled with a letter $\iota\notin T$ and a maximal event
$v_{\varepsilon}$ which is labeled with a letter $\varepsilon \notin T$.  We
notice that item \ref{definition:Process}.\ref{process:itemTwo} implies that
every condition which is not connected to an event $v\in V$ labeled by a
transition $t\in T$, is necessarily connected to $v_{\varepsilon}$.
Intuitively, $\iota$ loads the initial marking of $N$ and $\varepsilon$ empties
the marking of $N$ after the occurrence of all events of the process. We call
attention to the fact that the number of conditions connected to
$v_{\varepsilon}$ varies according to the process. 

\begin{figure}[hf] \centering \includegraphics[scale=0.40]{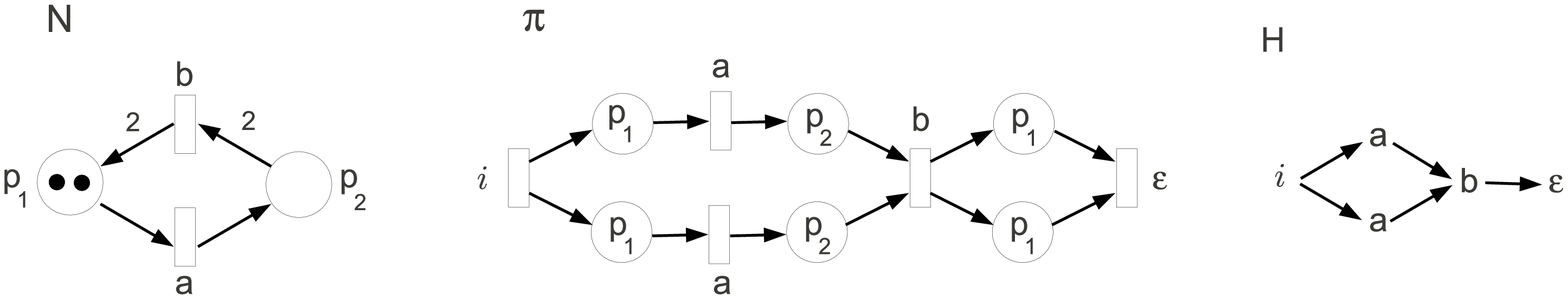}
\caption{ A $p/t$-net $N$, a process $\pi$ of $N$ and the Hasse diagram of the causal order
induced by $\pi$.}
\label{figure:process}
\end{figure}

A sequentialization of a partial order $po=(V,<,l)$ is another partial
order $po'=(V,<',l)$ satisfying $<\subseteq <'$. 
The causal order of a process $\pi$ is obtained from it by abstracting its
conditions and by considering the partial order induced by its events. An
execution is a sequentialization of a causal order. 

\begin{definition}[Causal Orders and Executions of $p/t$-net Processes]
\label{definition:CausalOrderExecution} The causal order of a process
$\pi=(B\dot{\cup}V,F,\rho)$ of a $p/t$-net $N$ is the partial order
$po_{\pi}=(V, <, l)$ where $ < = F^{*}|_{V\times V}$ and $l = \rho|_V$.  An
execution of $\pi$ is a sequentialization of $po_{\pi}$.  \end{definition}

We denote $\lang_{cau}(N)$ the set of all causal orders derived from processes
of $N$, $\lang_{ex}(N)$ the set of all its executions, and write simply $\lang_{PO}(N)$
whenever it is not relevant whether we are representing the set of causal orders
or the set of executions of $N$.  

\subsection{Interlaced Flows, Executions and Causal Orders} 
\label{section:InterlacedFlow}
\vspace{10pt}

Let $N=(P,T)$ be a $p/t$-net, $H=(V,E,l)$ a Hasse diagram with $l:V\rightarrow T$, 
and $p\in P$ be a place of $N$. Then a $p$-interlaced flow on 
$H$ with respect to $N$ is a four tuple $f=(\bb,\byf,\pb,\pf)$ of functions of 
type $E\rightarrow \N$ whose components satisfy the three following 
equations around each vertex $v$ of $H$: 
\vspace{5pt}
\begin{equation}
\label{equation:conservative}
\sum_{e^t=v} \byf(e) + \pf(e) = \sum_{e^s=v}\pb(e)+\pf(e)
\end{equation}
\begin{equation}
\label{equation:IN}
\mbox{\hspace{0.9cm}} In(v)=\sum_{e^t=v} \bb(e) + \pb(e) = \takesp(l(v))
\end{equation}
\begin{equation} 
\label{equation:OUT}
\mbox{\hspace{0.6cm}} Out(v)=\sum_{e^s=v} \bb(e) + \byf(e) = \putsp(l(v))
\end{equation}
\vspace{5pt}

Intuitively, for each $e\in E$, $\pb(e)$ counts some of the tokens produced in the  {\bf p}ast of $e^s$ and consumed {\bf b}y $e^t$;
$\pf(e)$, some of the tokens produced in the {\bf p}ast of $e^s$ and consumed in the {\bf f}uture of $e^t$, and $\byf(e)$, 
some of the tokens produced {\bf b}y $e^s$ and consumed in the future of $e^t$. Thus equation \ref{equation:conservative} states that on 
interlaced flows, the total number of tokens produced in the past of a vertex $v$, that arrives at it without 
being consumed, will eventually be consumed in the future of $v$. 
The component $\bb(e)$, counts the total number of tokens produced 
{\bf by} $e^s$ and consumed {\bf by} $e^t$. Thus, equation \ref{equation:IN} states 
that the total number of tokens consumed by $v$ is equal to $\takesp(l(v))$
while equation $\ref{equation:OUT}$ states that the total number of tokens 
produced by $v$ is $\putsp(l(v))$. Interlaced flows were introduced 
in \cite{deOliveiraOliveira2010} to characterize Hasse diagrams
of executions and causal orders of $p/t$-nets. This characterization is
formalized below in Theorem \ref{theorem:InterlacedFlow}. Intuitively 
it says that a Hasse diagram $H$ induces an execution of a given $p/t$-net $N$, 
if and only if it can be associated to a set of $p$-interlaced flows, one
for each place $p$ of $N$. A similar result holds with respect to Hasse diagrams of causal orders of $N$. The only difference 
is that if an edge $e$ belongs to the Hasse diagram of a causal order of $N$, then it must 
have arisen from a token that was transmitted from the event that labels its source vertex to the 
event that labels its target vertex, by using some place $p\in P$ as a channel. 
Thus in the flow that corresponds to $p$, the component  
which is responsible for the direct transmission of tokens must be strictly greater than zero. 

\begin{theorem}[Interlaced Flow Theorem\cite{deOliveiraOliveira2010}]
\label{theorem:InterlacedFlow}
Let $N=(P,T)$ be a (not necessarily bounded) $p/t$-net and $H=(V,E,l)$ be a Hasse diagram. Then 
\begin{itemize}
	\item[$(i)$] The partial order induced by $H$ is an execution of $N$ iff there exists 
		a $p$-interlaced flow $f_p:E\rightarrow \N^{4}$ in $H$ for each place $p$. 
	\item[$(ii)$] The partial order induced by $H$ is a causal order of $N$ iff 
	 there exists a set  $\{f_p\}_{p\in P}$ of $p$-interlaced flows such that for every 
	edge $e$ of $H$, the component $\bb_p(e)$ of $f_p(e)$, which denotes the direct transmission of
	tokens, is strictly greater than zero for at least one $p\in P$.
\end{itemize}
\end{theorem}

By using Theorem \ref{theorem:InterlacedFlow} we are able to provide a sliced characterization of executions 
and causal orders of $p/t$-nets. Namely, let $N=(P,T)$ be a $p/t$-net, $p\in P$ be a place of 
$N$ and $\boldS_1\boldS_2...,\boldS_n$ be a unit decomposition of a Hasse diagram $H$ such that 
$\boldS_i=(\{v_i\},E_i,l_i)$. Then a $p$-flow coloring of $\boldS_1\boldS_2...\boldS_n$ is a sequence 
of functions $f_p^1f_p^2...f_p^n$ with $f_p^i:E_i\rightarrow \N$ such that for any two consecutive 
slices $\boldS_i\boldS_{i+1}$, it holds that the value associated by $f_p^i$ to each edge $e$ 
touching the out frontier of $\boldS_i$ is equal to the value associated by $f_p^{i+1}$ to its 
corresponding edge touching the in-frontier of $\boldS_{i+1}$. We notice that a Hasse diagram 
$H$ has a $p$-interlaced flow $f_p$ with respect to $N$ if and only if each unit decomposition 
$\boldS_1\boldS_2...\boldS_n$ of $H=(V,E,l)$ admits a $p$-flow coloring $f_p^1f_p^2...f_p^n$. 
To see this, for each $e\in \boldS_i$ that is the sliced part of an edge $e'\in H$, set $f_p(e)=f_p(e')$.
In this way it makes sense to say that each $f_p^i$ is a {\em sliced $p$-interlaced flow} for $\boldS_i$. 

Now an {\em execution coloring} of a unit decomposition $\boldS_1\boldS_2...\boldS_n$ of a Hasse diagram 
with respect to a $p/t$-net $N$ is a sequence $F_1F_2...F_n$ where each $F_i=\{f_p^i\}_{p\in P}$ 
is a set of sliced $p$-interlaced flows for $\boldS_i$ where for each $i$ with $1\leq i\leq n-1$ 
and each $p\in P$, the value associated by $f_p^i$ to each edge $e$ touching the out-frontier of 
$\boldS_i$ is equal to the value associated by $f_p^{i+1}$ to its corresponding edge $e'$ touching 
the in-frontier of $\boldS_{i+1}$. A {\em causal coloring} of $\boldS_1\boldS_2...\boldS_n$, is 
an execution coloring $F_1F_2...F_n$ with the additional requirement that for each $i$, and 
each edge $e$ in $\boldS_i$, there is a $p\in P$ such that the component $\bb_{p}^i$ of $f_p^i\in F_i$
accounting for the direct transmission of tokes is strictly greater than $0$. Using the same 
argument as above we have that $\boldS_1\boldS_2...\boldS_n$ is a unit decomposition of an
execution (causal order) $H$ of $N$ if and only if it admits an execution (causal) coloring $F_1F_2...F_n$. 
We call each $F_i$, a {\em sliced execution-flow} ({\em sliced causal-flow}) for $\boldS_i$. The following 
proposition will be important for our refined characterization. 

\begin{proposition}
\label{proposition:SlicedInterlaced}
Let $N=(P,T)$ be a $k$-bounded $p/t$-net, $H$ be the Hasse diagram of an execution of $N$,
$\boldS_1\boldS_2...\boldS_n$ be a unit decomposition of $H$ where $\boldS_i=(\{v_i\},E_i,l_i)$, 
and $m_i:P\rightarrow \N$ be the marking of $N$ after the firing of the transitions 
$l(v_1)l(v_2)...l(v_i)$. Then 
 \begin{enumerate}[(i)]
	\item if $F_1F_2...F_n$ is an execution coloring of $\boldS_1\boldS_2...\boldS_n$ 
		then for each $f_p^i:E_i\rightarrow \N^4 \in F_i$ with $f_p^i(e)=(\bb_p^i(e),\byf_p^i(e),\pb_p^i(e),\pf_p^i(e))$ for $e\in E_i$, the following equation is satisfied
		\begin{equation}
		\label{equation:FlowOutFrontier}
			\sum_{e'} \bb_p^i(e') + \byf_p^i(e') + \pb_p^i(e') + \pf_p^i(e') =  m_i(p)
		\end{equation}
		where the sum is over all edges $e'$ touching the out-frontier of $\boldS_i$.
	\item if $F_1F_2...F_n$ is a causal coloring of $F$ then for each $i$,
		the size of the out-frontier of $\boldS_i$ is at most $k|P|$.  
\end{enumerate}
\end{proposition}

Intuitively, Proposition \ref{proposition:SlicedInterlaced}.$i$ says that in 
in a sliced execution flow, for each place $p\in P$ the sum of all tokens attached 
to the edges of the out-frontier of each unit slice $\boldS_i$ is equal to the number of 
tokens at place $p$ after the execution of the firing sequence $l(v_1)l(v_2)...l(v_i)$, 
where for $1\leq j\leq i$, $v_j$ is the center vertex of $\boldS_j$. 
Also notice that in a $k$-bounded $p/t$-net with $|P|$ places at most $k|P|$ tokens may be present 
in the whole net after each firing sequence. Thus, Proposition \ref{proposition:SlicedInterlaced}.$ii$
follows from Proposition \ref{proposition:SlicedInterlaced}.$i$ together with the fact that 
in a causal coloring $F_1F_2...F_n$ for each $i$ and edge $e\in \boldS_i$, the component $\bb_{p}^i(e)$ must be 
strictly greater than $0$ for at least one $p\in P$.$\square$

\begin{theorem}[Refined Expressibility Theorem]
\label{theorem:RefinedExpressibility}
Let $N$ be a $k$-bounded $p/t$-net. Then 
\begin{enumerate}[(i)]
	\item For any $c\geq 1$ there exists a (not necessarily saturated) Hasse diagram generator 
		$\hasseGenerator_{ex}^{c\exists}$ over $\slicealphabet^c$ representing 
		all executions of $N$ of existential slice width at most $c$. 
	\item For any $c\geq 1$ there exists a canonical saturated Hasse diagram generator $\hasseGenerator_{ex}^{c}$
		over $\slicealphabet^c$ representing all executions of $N$ of  global slice width at most $c$. 
	\item for any $c\geq 1$ there exists a canonical saturated Hasse diagram generator $\hasseGenerator_{cau}^c$ 
		over $\slicealphabet^c$ representing all the causal orders of $N$ of global slice width at most $c$.
		Furthermore, for any $c\geq k|P|$, we have that $\lang_{PO}(\hasseGenerator_{cau}^c) = \lang_{PO}(\hasseGenerator_{cau}^{k|P|})$. 
\end{enumerate}
\end{theorem}
\begin{proof}
$(i)$ Let $\slicegraph=(\mathcal{V},\mathcal{E},\mathcal{S})$ be a slice graph over $\slicealphabet^c$
where for each unit slice $\boldS\in \slicealphabet^c$, we have a vertex $\slicegraphvertex_{\boldS} \in \mathcal{V}$
with $\mathcal{S}(\slicegraphvertex) = \boldS$, and where there is an edge $(\slicegraphvertex_{\boldS},\slicegraphvertex_{\boldS'})$
if and only if $\boldS$ can be glued to $\boldS'$. Then clearly, a graph $G$ is in $\lang_{G}(\slicegraph)$ if and 
only if it has existential slice width at most $c$. Also let $\hasseGenerator=(\mathcal{V}',\mathcal{E}',\mathcal{S}')$ be the 
transitive reduced version of $\slicegraph$. At this point we have that a partial order $po$ is in $\lang_{PO}(\hasseGenerator)$
if and only if its Hasse diagram has existential slice width $c$. Now we will describe how to filter out from $\lang_{PO}(\hasseGenerator)$
all the partial orders that are not executions of $N$: Define the Hasse diagram generator
$filter_{ex,c}^N(\hasseGenerator)=(\mathcal{V}'',\mathcal{E}'',\mathcal{S}'')$ as follows: For each vertex $\slicegraphvertex_{\boldS} \in \mathcal{V}$
and each sliced execution flow $F=\{f_p\}_{p\in P}$ of $\boldS$ such that for each $p\in P$, $\sum_{e}f_p(e)\leq c$ where $e$ ranges over 
the edges touching the out-frontier of $\boldS$, we add a vertex $\slicegraphvertex_{\boldS,F}$ to $\mathcal{V}''$ and label this vertex with $\boldS$. Furthermore, 
we add an edge $(\slicegraphvertex_{\boldS,F} , \slicegraphvertex_{\boldS',F'})$ to $\mathcal{E}''$ if and only if 
$(\slicegraphvertex_{\boldS},\slicegraphvertex_{\boldS'}) \in \mathcal{E}$ and if $(\slicegraphvertex,F)$ can be glued to $(\slicegraphvertex',F')$. 
In this way, there is a path $\slicegraphvertex_{\boldS_1,F_1}\slicegraphvertex_{\boldS_2,F_2}...\slicegraphvertex_{\boldS_n,F_n}$ from an 
initial to a final vertex in $filter_{ex,c}^N(\hasseGenerator)$ if and only if $F_1F_2...F_n$ is an execution coloring of $\boldS_1\boldS_2...\boldS_n$. By Theorem 
\ref{theorem:InterlacedFlow}.$i$ such a coloring exists if and only if the Hasse diagram $H= \boldS_1\circ \boldS_2 \circ ... \circ \boldS_n$ has existential 
slice width at most $c$ and if its induced partial order is an execution of $N$. Therefore we can set $\hasseGenerator_{ex}^{c\exists}(N)$ to be 
$filter_{ex,c}^N(\hasseGenerator)$.

$(ii)$ Let $\hasseGenerator^c$ be the Hasse diagram generator of Lemma \ref{lemma:UniversalHasseDiagramGenerator} whose 
graph language consists precisely of the Hasse diagrams of global slice width at most $c$. Now let $\hasseGenerator^{\cap}$ be the 
Hasse diagram generator whose graph language consists in the intersection  $\lang_G(filter_{ex,c}^N(\hasseGenerator))\cap \lang_{G}(\hasseGenerator^c)$. 
By Lemma \ref{lemma:ClosuresGraphLanguage}.$1$, $\hasseGenerator^{\cap}$ can be effectively constructed from 
$\hasseGenerator^c$ and from $filter_{ex,c}^N(\hasseGenerator)$, observe 
that since $filter_{ex,c}^N(\hasseGenerator)$ is not saturated, it is not evident that $\hasseGenerator^{\cap}$ is saturated. Nevertheless, 
the saturability of $\hasseGenerator^{\cap}$ follows from Theorem \ref{theorem:InterlacedFlow}.$i$ and from the fact that whenever 
a set $F=\{f_p\}_{p\in P}$ of $p$-interlaced flows can be associated to a Hasse diagram $H$, we have that any unit decomposition $\boldS_1\boldS_2...\boldS_n$
of $H$ admit an execution coloring $F_1F_2...F_n$, where each $F_i$ is the sliced part of $F$ that is associated to $\boldS_i$. Since $\hasseGenerator^{\cap}$ is 
saturated, by Theorem \ref{theorem:ClosuresPartialOrderLanguage}.$3$ there is a canonical Hasse diagram generator 
$\hasseGenerator_{ex}^c = \mathcal{C}(\hasseGenerator)$ representing the same graph language $\hasseGenerator^{\cap}$, and consequently 
the same partial order language. 

$(iii)$ The proof is analogous to the proof of item $(ii)$, except for two small adaptations: Replace each occurrence of the word "execution" by 
the word "causal", and each occurrence of the subscript $ex$ by the subscript $cau$. In this way whenever $F_1F_2...F_n$ appears in the proof 
it will denote a causal coloring instead of an execution coloring. Analogously the filter $filter_{cau,c}^N(\hasseGenerator)$ will filter out 
from the graph language of $\hasseGenerator$ all the Hasse diagrams whose induced partial order is not a causal order of $N$. The only 
additional caveat, is that since the Hasse diagram of any causal order of $N$ has global slice width at most $k|P|$, we have that 
$\lang_{PO}(\hasseGenerator_{cau}^c) = \lang_{PO}(\hasseGenerator_{cau}^{k|P|})$ whenever $c\geq k|P|$. Therefore the whole 
causal order behavior of $N$ can be represented by $\hasseGenerator_{cau}^{k|P|}$.  

\end{proof}

\begin{figure}[h]
\centering
\includegraphics[scale=0.35]{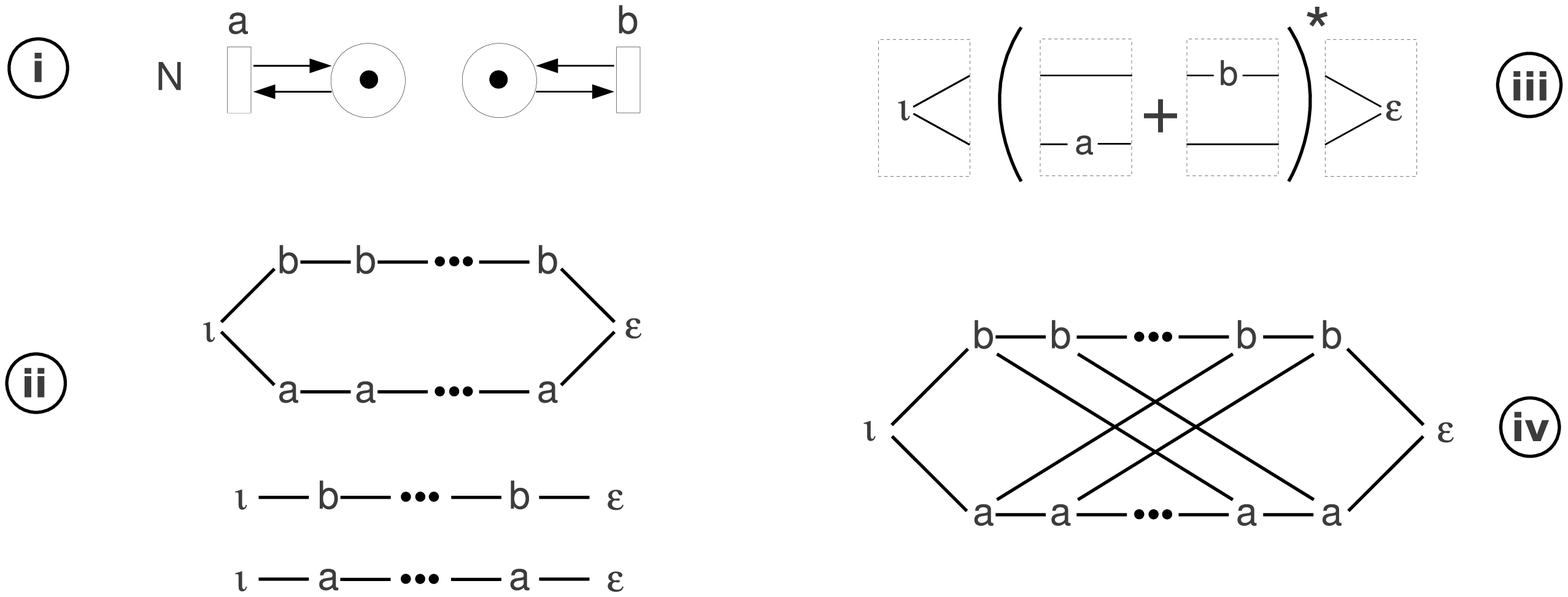}
\caption{ $i$) A $1$-bounded $p/t$-net $N$ $ii$) The causal language $\lang_{cau}(N)$
of $N$
$iii$) A weakly saturated (but not yet transitive reduced) slice language representing 
$\lang_{cau}(N)$ $iv$) An intuitive depiction of a family of partial orders that 
is included in the set $\lang_{ex}(N)$ of sequentializations of partial orders in $\lang_{cau}(N)$, but that is not 
representable trough slice languages (not even non-saturated slice languages)
over a finite slice alphabet.}
\label{figure:CausalExecution}
\end{figure}

We observe it may be the case that the graph language of $\lang_{G}(\hasseGenerator_{ex}^c(N)) \subseteq \lang_{G}(\hasseGenerator_{ex}^{c+1}(N))$ 
for every $c+1$. And indeed this fact can already be noticed in the execution behavior of rather simple nets such as the one depicted in Figure \ref{figure:CausalExecution}.$i$.
While the causal behavior $\lang_{cau}(N)$ of $N$, which is intuitively depicted in Figure \ref{figure:CausalExecution}, is relatively simple, 
and can be easily described in terms of regular slice languages (Figure \ref{figure:CausalExecution}.$iii$), the set 
$\lang_{ex}(N)$ of all sequentializations of partial orders in $\lang_{cau}(N)$, contains subfamilies of partial orders
whose Hasse diagrams have unbounded existential slice width, and that for this reason, 
cannot be represented through slice languages over finite alphabets. An example of such a subfamily 
is the set of partial orders induced by the sequence $\{H_n=(V_n,E_n,l_n)\}_{n\in \N}$ of Hasse diagrams defined below and 
depicted in Figure \ref{figure:CausalExecution}.$iv$.

\begin{equation*}
V_n=\{v_{\iota},v_{\varepsilon}\} \cup \{v_{a_1},...,v_{a_{n}}\}\cup \{v_{b_1}, ...,v_{b_n}\}
\end{equation*}
\begin{equation*}
l(v_{\iota})= \iota \mbox{,   } l(v_{\varepsilon}) = \varepsilon \mbox{,  } l(v_{a_i}) = a \mbox{,  } l(v_{b_i})=b
\end{equation*}
\begin{equation*}
\begin{array}{rcl}
E_n & = &\{(v_{\iota},v_{a_1}),(v_{\iota},v_{b_1}),(v_{a_n},v_{\varepsilon}),(v_{b_n},v_{\varepsilon})\}\hspace{0.1cm} \cup \\
& & \{(v_{a_i},v_{a_{i+1}})\} \cup \{(v_{b_i},v_{b_{i+1}})\} \hspace{0.1cm} \cup \\
& & \{(v_{a_i},v_{b_{n-i}})\} \cup \{(v_{b_{i}},v_{a_{n-i}})\}
\end{array}
\end{equation*}

Therefore, 
the parametrization of the language of executions of a $p/t$-net with respect to the maximal global slice 
width of the respective Hasse diagrams is unavoidable, if we are willing to represent executions via 
regular slice languages. When dealing with the causal behavior of bounded $p/t$-nets such a 
parametrization is not essential since the whole causal behavior of $k$-bounded $p/t$-nets with set of places $P$
can already be captured by regular slice languages over $\slicealphabet^{k|P|}$. In particular, 
a neat implication of Theorem \ref{theorem:RefinedExpressibility}.$iii$ is that if a $k$-bounded $p/t$-net $N=(P,T)$ 
and a $k'$-bounded $p/t$-net $N'=(P',T)$ have the same partial order behavior, then $\hasseGenerator_{cau}^{k|P|}(N)\simeq \hasseGenerator_{cau}^{k'|P'|}(N')$. 

\begin{theorem}[$p/t$-nets and Hasse diagram generators \cite{deOliveiraOliveira2010}]
\label{theorem:HasseDiagramGeneratorsPetriNets} \mbox{}\\ 
Let $\lang_{PO}$ be a partial order language generated by a (not necessarily saturated) Hasse diagram generator $\hasseGenerator$ 
over $\slicealphabet^c$.
\begin{itemize} 
\item {\bf Verification: }\label{theorem:Verification} Let $N$
be a bounded $p/t$-net. Then the following problems are decidable:
	\begin{itemize}
		\item Is $\lang_{ex}(N) \cap \lang_{PO} = \emptyset$ ? 
		(Is $\lang_{cau}(N) \cap \lang_{PO} = \emptyset$ ?)
		\item Is $\lang_{PO} \subseteq \lang_{ex}(N)$? 
			(Is $\lang_{PO} \subseteq \lang_{cau}(N)$?)
	\end{itemize}
\item {\bf Synthesis: }\label{theorem:Synthesis} Let $k\geq 1$ and $r\geq 1$. Then it is
possible to determine if it exists, and if so, automatically synthesize
\begin{itemize}
	\item a $k$-bounded $p/t$-net $N$ whose execution behavior $\lang_{ex}(N)$ minimally includes $\lang_{PO} $. 
	\item a $k$-bounded $p/t$-net $N$ with place repetition number $r$ whose causal behavior $\lang_{cau}(N)$
		minimally includes $\lang_{PO}$.  
\end{itemize}
\end{itemize}
\end{theorem}

We point out that in order to carry the verification result stated in Theorem \ref{theorem:HasseDiagramGeneratorsPetriNets} 
we do not need to construct the Hasse diagram generators $\hasseGenerator_{ex}^{c\exists}$,$\hasseGenerator_{ex}^c$ and $\hasseGenerator_{cau}^c$. 
Instead we may apply the filters used in the proof of Theorem \ref{theorem:RefinedExpressibility} directly to the 
Hasse diagram generator we want to verify. In what follows let $sem=ex$ if we are performing the verification according to the 
execution semantics and $sem=cau$ if we are performing the verification according to the causal semantics. Then we 
have that $\lang_{sem}(N)\cap \lang_{PO}(\hasseGenerator) =\emptyset$ if and only if the slice language of the 
filtered version of $\hasseGenerator$ is empty, i.e., if $\lang(filter_{sem,c}^{N}(\hasseGenerator)) = \emptyset$, 
while $\lang_{PO}(\hasseGenerator)\subseteq \lang_{ex}(N)$ if and only if the slice language of the filtered version of 
$\hasseGenerator$ equals the slice language of $\hasseGenerator$ itself, i.e, if $\lang(filter_{sem,c}^{N}(\hasseGenerator)) = \lang(\hasseGenerator)$.
Since the slice languages represented by Hasse diagram generators are regular, and since emptiness and equality are decidable 
for regular languages, the verification results hold. 

Concerning the synthesis with respect to the causal semantics we need to specify \`{a} priori, the maximum number $r$
of repeated copies a place is allowed to have in the synthesized net. This additional parameter is not necessary when considering the 
synthesis with respect to the execution semantics because the set of executions of a $p/t$-net remains invariant upon the addition of a 
place that is already part from the net. Adding a repeated place to a $p/t$-net may however increase its causal behavior
by increasing the possibility of causal interactions between its transitions. We refer to \cite{deOliveiraOliveira2010} for a detailed discussion on this topic. 

We finish this section by stating a corollary  (Corollary \ref{corollary:MSCAndPetriNets}) that extends the applicability 
of slice graphs and Hasse diagram generators by showing that they can serve as an interface 
between $p/t$-nets and other well known formalisms for the specification of partial order languages, such as 
$MSC$-languages and Mazurkiewicz trace languages. More precisely, it states that both the 
{\em verification} and the {\em synthesis} results of Theorem \ref{theorem:HasseDiagramGeneratorsPetriNets}
can be reformulated in terms of these formalisms. Indeed, as we showed in Section \ref{section:Reductions} 
both formalisms can be mapped to slice graphs, which in general generate non transitive reduced
slice languages. By an application of our transitive reduction algorithm (Theorem \ref{theorem:TransitiveReductionSliceGraphsB})
these slice graphs can be transformed into Hasse diagram generators, with the aim to 
meet the requirements of Theorem \ref{theorem:HasseDiagramGeneratorsPetriNets} (which is not valid for general slice graphs).
Finally these Hasse diagram generators can be used to address both the verification and the 
synthesis of $p/t$-nets as stated in Theorem \ref{theorem:HasseDiagramGeneratorsPetriNets}. 

\begin{corollary}[MSC Languages, Mazurkiewicz Traces and $p/t$-nets]
\label{corollary:MSCAndPetriNets} 
The synthesis and verification results stated in Theorem \ref{theorem:HasseDiagramGeneratorsPetriNets}
is equally valid if the partial order language $\lang_{PO}$ is represented by a pair
$(\automaton,I)$ of finite automaton and independence relation, or by a message sequence graph $\mathcal{M}$. 
\end{corollary}

We emphasize that the verification and synthesis results of Theorem \ref{theorem:HasseDiagramGeneratorsPetriNets} 
do not require the Hasse diagrams to be saturated. Analogously, Corollary 
\ref{corollary:MSCAndPetriNets} does not require the language specified by the pair $(\automaton,I)$ to be recognizable 
(i.e. the trace closure of $\lang(A)$ to be recognized by a finite automaton), 
nor the partial order language specified by the message sequence graph $\mathcal{M}$ 
to be linearization-regular. For a matter of comparison we point out that the synthesis of labeled $p/t$-nets 
(i.e., nets in which two transitions may be labeled by the same action) from
recognizable Mazurkiewicz trace languages (and indeed more generally 
from the local trace languages \cite{HoogersKleijnThiagarajan1996})
was addressed in \cite{HussonMorin2000,KuskeMorin2002}. Corollary \ref{corollary:MSCAndPetriNets} 
concerns the synthesis of unlabeled $p/t$-nets. To contrast the partial order behavior
of labeled and unlabeled $p/t$-nets, we notice that labeled $1$-safe $p/t$-nets are
already as partial order expressive as their $b$-bounded counterparts
\cite{BestWimmel2000}, while this is not the case for unlabeled $p/t$-nets. 
Thus the synthesis of unlabeled nets which is addressed in Corollary \ref{corollary:MSCAndPetriNets} 
tends to be harder.

\section{Final Comments}
\label{section:FinalComments}

The main contributions of the present work were twofold. 
First, we devised an algorithm that transitive reduces any 
slice graph into a Hasse diagram generator
representing the same set of partial orders. 
Second, we developed the theory of saturated slice languages, 
which lifts some of the most intuitive aspects of trace 
theory to the slice setting.
From a conceptual perspective our transitive reduction algorithm conciliates
the flexibility of reasoning about partial orders in terms of DAGs,
which is implicit in most of the literature dedicated to the representation
of infinite families of partial orders, with the aesthetical and algorithmic 
advantages of specifying such families through sets of Hasse diagrams.
From a practical perspective our algorithm turned to be a necessary step towards 
putting distinct concurrency theoretic formalisms such as Mazurkiewicz traces, 
message sequence charts and Petri nets, into a common ground with respect 
to the partial order languages they represent. As a consequence 
we were able to address the verification and automatic synthesis 
of concurrent systems from an unified perspective.
By combining our transitive reduction with our development of saturated 
slice languages we were able to address the canonization of 
slice graphs with respect to their partial order languages and to 
prove several decidability and computability results respective 
to the manipulation of these slice graphs. Furthermore we showed 
via reductions that all these results hold as well for partial order 
languages represented by recognizable Mazurkiewicz trace languages, 
by recognizable MSC languages specified by Message sequence graphs, and 
by Petri nets. Therefore we consider 
that the overall contribution of the present work consists in a robust
methodology to compare and operate with partial order languages generated by 
seemingly disconnected formalisms.

\bibliographystyle{abbrv}
\bibliography{canonizablePartialOrderGeneratorsArxivVersion4}

\end{document}